\documentclass[fleqn,usenatbib]{mnras}

\usepackage{newtxtext,newtxmath}

\usepackage[T1]{fontenc}

\DeclareRobustCommand{\VAN}[3]{#2}
\let\VANthebibliography\thebibliography
\def\thebibliography{\DeclareRobustCommand{\VAN}[3]{##3}\VANthebibliography}

\usepackage{graphicx}	
\usepackage{amsmath}	
\usepackage{multirow}
\usepackage[version=4]{mhchem}
\usepackage{subcaption}
\usepackage{hyperref}
\usepackage{threeparttable}
\usepackage[symbol]{footmisc}
\usepackage{pdflscape}
\usepackage{afterpage}
\usepackage{makecell}

\newcommand{\minitab}[2][@{}l]{\begin{tabular}{#1}#2\end{tabular}}

\title[Heavily polluted cool DAZs]{I Can Do It With A Broken Planetesimal: Characterising the planetary debris in heavily polluted cool white dwarfs}

\author[A.-M. Cutolo et al.]{Anna-Maria Cutolo,$^{1}$\thanks{E-mail: anna-maria.cutolo@warwick.ac.uk}
Boris T. G\"ansicke,$^{1}$
Andrew Swan,$^{1}$
Andrew M. Buchan,$^{1}$
Jamie T. Williams,$^{1}$
\newauthor Detlev Koester,$^{2}$
Paula Izquierdo,$^{1}$
Mark A. Hollands,$^{1}$
James A. Blake,$^{1}$
Nicola P. Gentile Fusillo$^{3}$
\\
$^{1}$Department of Physics, University of Warwick, Coventry, CV4 7AL, UK\\
$^{2}$Institut f\"ur Theoretische Physik und Astrophysik, University of Kiel, 24098 Kiel, Germany\\
$^{3}$Universita degli studi di Trieste, Via Valerio, 2, Trieste, 34127, Italy\\
}

\date{Accepted XXX. Received YYY; in original form ZZZ}

\pubyear{2026}

\begin{document}
\label{firstpage}
\pagerange{\pageref{firstpage}--\pageref{lastpage}}
\maketitle

\begin{abstract}
We present the analysis of four cool and strongly metal-polluted H-atmosphere white dwarfs observed with X-shooter. We  compared their atmospheric parameters obtained from spectroscopy, photometry and a hybrid method, finding a difference of up to $\simeq160$\,K in $T_\mathrm{eff}$, and $\simeq0.3$\,dex in $\log g$, between the three analyses. We adopt the $T_\mathrm{eff}$ and $\log g$ from the photometric analysis, and subsequently measured the metal abundances of the photospheric elements from spectroscopic modelling, analysing their compositions. We identified from five to eleven unique metals in the photospheres of the four white dwarfs, with total accretion rates ranging from $10^{8}$ to $10^{9}$\,$\mathrm{g~s^{-1}}$. The compositional analysis of WD\,J035826.49+215726.16 suggests an accreted planetesimal akin to a core-rich differentiated body, with a core mass fraction of 70~per cent, placing it among the most core-rich objects known to be accreted by white dwarfs. The parent body accreted by WD\,J042643.98--415341.44 shows an enhancement in Na compared to that of the Earth, making it most similar to primitive chondrites. The photosphere of WD\,J013222.88+052923.71 is greatly depleted in core-like material, and its composition resembles that of pure crust/mantle material. WD\,J232428.21--021643.65 has accreted the most Fe-rich planetesimal, with an Fe mass fraction of 67~per cent. We compare these results to other published studies and conclude that these white dwarfs are among the most heavily polluted cool DAZs studied to date, increasing the sample of cool H-dominated white dwarfs with five or more metals by 50~per cent.
\end{abstract}

\begin{keywords}
planets and satellites: composition -- stars: abundances -- stars: white dwarfs
\end{keywords}



\section{Introduction}
White dwarf planetary systems is a rapidly growing area of study in the exoplanet field of science \citep{zuckerman1987excess, farihi2016circumstellar, veras2016post, xu2024chemistry}. Almost all known low- to intermediate-mass stars hosting planets will evolve off the main sequence and ultimately end their lives as a white dwarf. Depending on the distance from the star, some planetary bodies will survive destruction in the giant branch phase \citep{villaver2007can, mustill2012foretellings}. These bodies can be tidally disrupted if they pass within the Roche radius of the star, and this material can subsequently be accreted onto the white dwarf \citep{jura2003tidally, veras2014formation}. The observational evidence for these planetary remnants can come in five forms: infrared excess from circumstellar dust (e.g. \citealt{graham1990infrared, favieres2024sample}), double-peaked emission lines from gas discs (e.g. \citealt{gaensicke2006gaseous, manser2016another}), transits from disintegrating bodies (e.g. \citealt{vanderburg2015disintegrating, guidry2025transiting}), X-rays from accreting debris \citep{cunningham2022white}, and metal pollution in the atmosphere of the white dwarf (e.g. \citealt{koester1997metals, klein2010chemical, raddi2015likely, wilson2015composition, xu2017chemical, hoskin2020white, williams2025measurements})~--~the latter of which we will be analysing in this paper.

White dwarfs are extremely dense objects for their size, supported against gravitational collapse by electron degeneracy pressure, which elicits far higher surface gravity than main sequence stars. Due to this high surface gravity, the atmospheres are primarily composed of H or He, with metals sinking  below their atmospheres on timescales that are orders of magnitude shorter than their cooling age \citep{schatzman1948spectrum, koester2009accretion}. However, white dwarfs with an effective temperature $\gtrsim20\,000$\,K are hot enough to sustain selective metals in their photosphere through a process known as radiative levitation, which can cloud the interpretation of debris composition \citep{chayer1995radiative}. Below this temperature, metal absorption features that appear in the spectra are a clear sign of the recent or ongoing accretion of planetary debris onto the white dwarf (see e.g. \citealt{zuckerman2011aluminum, swan2019interpretation, izquierdo2021gd, aguilera2025host}). Although interstellar medium (ISM) accretion has been debated as a possible cause, data from \textit{Spitzer} and ground-based observations have proven that the source of heavy elements for polluted white dwarfs is in fact rocky exoplanetary material being accreted from a debris disc around the white dwarf \citep{farihi2010strengthening}.

Metal-polluted white dwarfs provide us with detailed compositional insights into the interior of exoplanets, which are not possible to determine when using traditional methods of exoplanet detection around main sequence stars. Transit and radial velocity methods can measure the radius and mass of exoplanets \citep{charbonneau1999detection, mandel2002analytic, seager2011exoplanets} and by fitting mass-radius relations, the interior can be inferred, although this method is subject to degeneracies \citep{dorn2015can}. However, by measuring the abundances of metals present in the photosphere of a white dwarf, we are able to deduce more detailed properties of the exoplanetary material that was accreted, such as whether it is accreting core-, mantle-, or crust-rich fragments of differentiated bodies (e.g. \citealt{zuckerman2011aluminum, melis2011accretion}) or undifferentiated material such as chondrites (e.g. \citealt{izquierdo2021gd, doyle2023new}). Fe, Mg, and Ca abundances can be used as core, mantle, and crust indicators, respectively \citep{hollands2018cool}, and the detailed composition analysis is generally compared against the Solar System benchmarks of bulk Earth, its differentiated components, and meteorite families.

It has been estimated that 20-50~per cent of white dwarfs are metal polluted (e.g. \citealt{zuckerman2003metal, koester2014frequency, rouis2024constraints, manser2024frequency}), although most that have been classified as such only exhibit one or two heavy elements in their spectra ($\sim$75~per cent, the strongest lines usually being Ca H~\&~K resonance lines \citealt{williams2024pewdd}), which is insufficient for detailed analysis of the debris composition \citep{dufour2007spectral}. There have, however, been several detailed abundance studies, primarily of He-atmosphere polluted white dwarfs (e.g. \citealt{klein2010chemical, hoskin2020white, izquierdo2021gd, doyle2023new}). Metals are more visible in He-atmosphere white dwarfs because He opacity is lower than that of H. Also, the sinking timescales of metals are orders of magnitude longer in He-atmosphere stars ($10^6$\,yr) than in their H counterparts, which can be as short as days in the hotter stars. 

White dwarf discs are only estimated to exist for $10^4-10^6$\,yr \citep{girven2012constraints, veras2020lifetimes, cunningham2021horizontal}, and thus the accretion episode might have ended~--~with heavier metals already sinking out of the atmosphere~--~when the He-atmosphere white dwarfs are observed. This can skew the analysis, as it not possible to predict how long ago accretion ended and hence account for the metal diffusion that would have occurred. Therefore, in these cases the true composition of the accreted material is difficult to determine. 

On the contrary, the shorter sinking timescales in a H-atmosphere white dwarf mean that it is highly likely to have been observed during a period of ongoing accretion in which the abundances can be directly traced back to the accreted body, making these highly useful for compositional analysis. There are fewer detailed abundance studies of these, and even fewer for the cool ($<8000$\,K) ones due to observational bias: H atmospheres have a higher opacity at these temperatures and so the equivalent widths of metal lines are about a factor of 1000 smaller than in their He counterparts \citep{dupuis1993study, zuckerman2003metal}. We will refer to polluted cool H-atmosphere white dwarfs as DAZs and polluted cool He-atmosphere white dwarfs as DZs, following convention (where A denotes dominant H lines and Z denotes metal lines in the spectra).

In this paper, we present a detailed analysis of four cool DAZ white dwarfs which show some of the highest metal abundances in their temperature range. WD\,J035826.49+215726.16 \citep[WD\,J0358, ][]{tremblay2020gaia} and WD\,J042643.98--415341.44 \citep[WD\,J0426, ][]{obrien2023gaia} were identified as being heavily polluted when observed to complete the 40\,pc volume-limited sample \citep{{obrien202440}}.  WD\,J013222.88+052923.71 \citep[WD\,J0132, ][]{gentile2015photometric} and WD\,J232428.21--021643.65 \citep[WD\,J2324, ][]{gentile2015photometric} exhibited strong Ca absorption in SDSS spectra \citep{kepler2015new, blouin2022no}. All four were followed up using X-shooter, the results of which are presented here.

\section{Observations}
\label{sec:data}
\begin{figure*}
    \includegraphics[width=\textwidth]{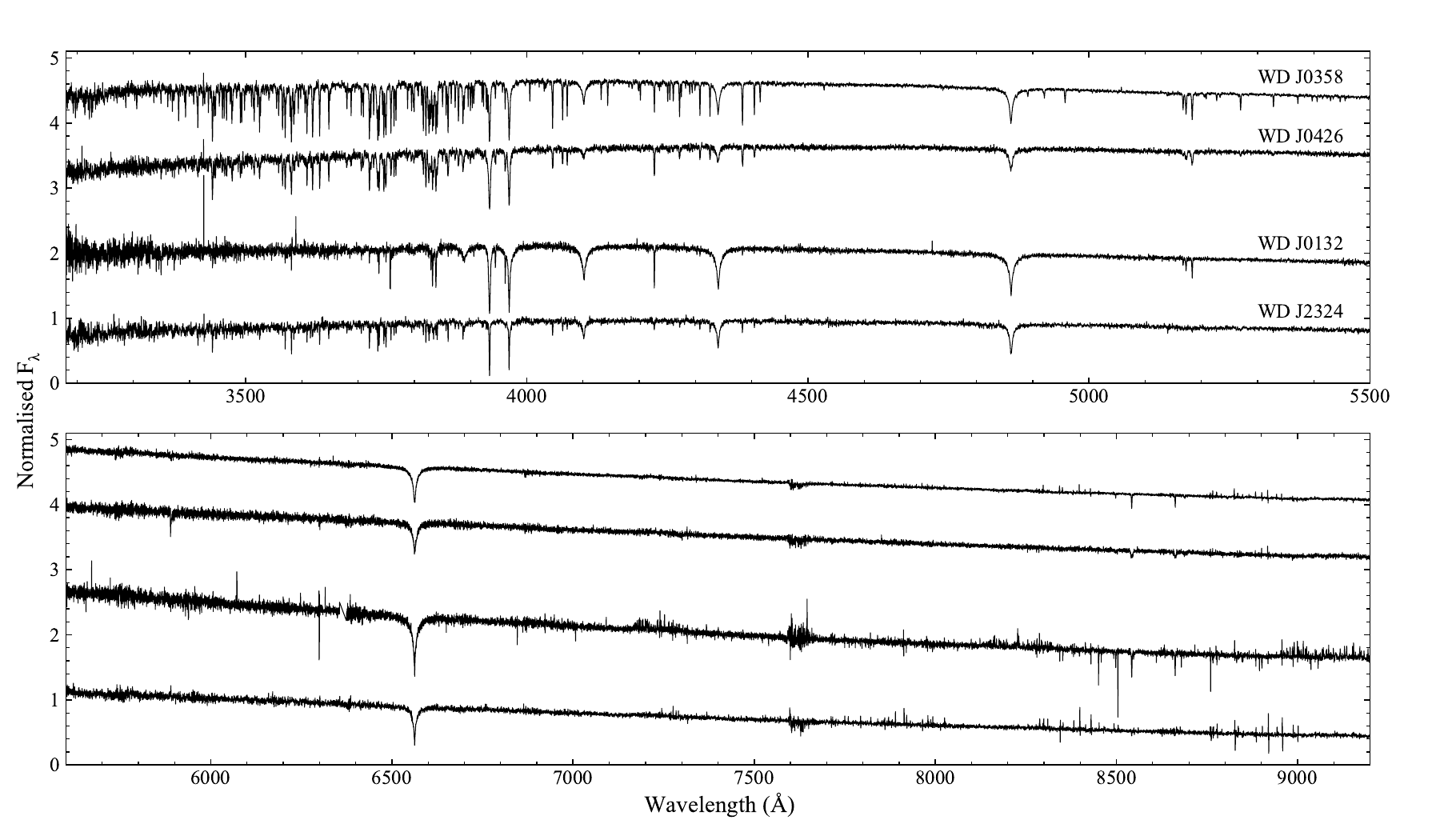}
    \caption{X-shooter spectra of the four DAZs. Top panel shows the UVB spectra and bottom panel shows the VIS spectra. The spectra are normalised using \hyperref[eq:1]{Eq.\,\ref*{eq:1}} and vertically offset from unity for visibility.}
    \label{fig:xshooter_spectra}
\end{figure*}

The spectroscopy analysed in this paper was obtained using the X-shooter echelle spectrograph \citep{vernet2011x} on the Very Large Telescope at Cerro Paranal, with the observations detailed in \autoref{tab:observations}. This instrument provides simultaneous optical and near-infrared coverage across three independent arms. Wavelength coverages are 2989$-$5560\,\AA, 5337$-$10\,200\,\AA\ and 9940$-$21\,010\,\AA\ in the UVB, VIS, and NIR arms, respectively. Slit widths were set to 1.0, 0.9, and 0.9\,arcsec, for nominal resolving powers $R=\lambda/\Delta\lambda$ of 5453, 8935, and 5567 for the UVB, VIS, and NIR arms, respectively. The observations were taken in STARE mode and the data were reduced following a standard procedure employing the \textsc{EsoReflex} pipeline \citep{freudling2013automated}. The reduced spectra were corrected for telluric absorption using \textsc{molecfit} \citep{smette2015molecfit, kausch2015molecfit}. The flux calibration used observations of DA white dwarfs obtained with the same instrument setup as the science spectroscopy. The data from the NIR arm had a low signal-to-noise ratio ($\mathrm{S/N}< 10$) and so were not included in the analysis presented here. The co-added X-shooter spectra for the four white dwarfs are shown in \hyperref[fig:xshooter_spectra]{Fig.\,\ref*{fig:xshooter_spectra}}. 

We complemented the X-shooter spectroscopy with photometric data spanning the ultraviolet (\textit{GALEX}), optical (\textit{Gaia}, Pan-STARRS, SDSS, and SkyMapper), and infrared (2MASS, UKIDSS, VISTA, and \textit{WISE}) (see \autoref{tab:phot_table}).

\begin{table}
    \centering
    \caption{Log of observations. The values correspond to the UVB and VIS observations, respectively, with the signal-to-noise ratio S/N quoted from the ESO archive.}
    \label{tab:observations}
    \begin{tabular}{lcccc}
	\hline
        Target & Total Exposure Time (s) & S/N & Date and Time\\
        \hline
        WD\,J0358 & 600, 550 & 33, 30 & 2019-10-07 06:45\\
        & & 31, 28 & 2019-10-07 06:56\\
        & & 19, 20 & 2019-10-08 05:46\\
        & & 21, 22 & 2019-10-08 05:57\\
        & & 23, 23 & 2019-10-08 06:11\\
        & & 23, 23 & 2019-10-08 06:22\\
        & & 24, 23 & 2019-10-08 06:36\\
        & & 28, 26 & 2019-10-08 06:47\\
        WD\,J0426 & 481, 431 & 28, 26 & 2019-10-07 07:13\\
        & & 28, 27 & 2019-10-07 07:21\\
        WD\,J0132 & 2950, 2840 & 26, 17 & 2014-08-04 08:00\\
        & & 33, 22 & 2014-08-26 07:09\\
        WD\,J2324 & 1250, 1220 & 11, 9 & 2018-07-11 05:36\\
        & & 15, 12 & 2018-07-11 05:58\\
        & & 18, 15 & 2018-07-11 06:23\\
        & & 14, 12 & 2018-07-11 06:45\\
        & & 16, 13 & 2018-07-11 07:10\\
	\hline
    \end{tabular}
\end{table}

\section{Atmospheric Analysis}
\label{sec:analysis}
\subsection{White Dwarf Parameter Determination}
\subsubsection{Overview}
In order to measure the photospheric metal abundances of the four white dwarfs, the white dwarf atmospheric parameters of effective temperature ($T_\mathrm{eff}$) and surface gravity ($\log g$) must be determined. This can be done through either spectroscopic (e.g. \citealt{koester2015db}) or photometric (e.g. \citealt{bergeron1997chemical}) methods; both requiring synthetic atmospheric model spectra. It has been shown that spectroscopic and photometric analyses generally arrive at different solutions (with larger differences for He-dominated atmospheres, see e.g. \citealt{genest2019comprehensive}). $T_\mathrm{eff}$ derived from only photometric data tend to be systematically lower with no general trend in $\log g$. The opacity of metals affects the structure of the atmospheres of strongly polluted white dwarfs, so it needs to be taken into account in the fit \citep{tremblay2019fundamental, izquierdo2023systematic, le2025tracing}. Simultaneous spectroscopic and photometric fitting (e.g. \citealt{izquierdo2021gd}) results in parameters that make use of the specific constraints provided by each method. We explore the different fitting methods and report the atmospheric parameters determined from each. Subsequently, the metal abundances are calculated from synthetic spectra set to the photometric parameters.

He has no strong transitions at the effective temperature of these white dwarfs. However, its presence can induce Balmer-line broadening in cool H-dominated white dwarfs \citep{bergeron1991synthetic, koester2005hs} but no white dwarfs in our sample show evidence of this. \citet{le2025tracing} find that even high He abundances ($\log(\mathrm{He/H})\lesssim-1$) do not result in significant changes in sinking timescales and hence we do not consider the presence of He.

\subsubsection{Model Atmospheres and Synthetic Spectra}
\label{sec:models}
We used the \cite{koester2010white} model atmosphere code to generate grids of synthetic spectra across the parameter space required for the analysis. In an initial step, we fitted the spectroscopy and photometry using a 2D DA grid to estimate $T_\mathrm{eff}$ and $\log g$. 

We then created 3D model grids for each star with $T_\mathrm{eff}$ and $\log g$, and included an overall metallicity defined in terms of the Si abundance, $\log(\mathrm{Si/H})$, as parameters. We included O, Na, Mg, Al, Si, Ca, Ti, Cr, Mn, Fe, Co, and Ni in the models, with their abundances set relative to Si at bulk Earth values \citep{mcdonough2000}. Other trace metals present in a bulk Earth composition neither have strong transitions in the optical, nor affect the atmospheric structure, and hence are not included in the model. Accounting for metals in the computation of the atmospheric structures is important because of their significant contribution to the opacity. These 3D grids typically spanned 1000\,K in $T_\mathrm{eff}$ using steps of 200\,K, 1\,dex in $\log g$ using steps of 0.25\,dex, and $\log(\mathrm{Si/H}) = -4.5$ to $-8.5$ in steps of 0.5\,dex.

\subsubsection{Fitting Method}
We fitted the grids of synthetic spectra to both the spectroscopic and photometric data using the Markov Chain Monte Carlo (MCMC) method. The MCMC routine used to determine the best-fitting parameters makes use of the Python package \textsc{emcee} \citep{foreman2013emcee}. 

The parameter space was explored using an iterative approach. Walkers can take time to find the high-probability regions which can lengthen the autocorrelation time and therefore the entire sampling run. We overcame this by using three passes with increasing walkers and steps to quickly explore the parameter space with short runs, with each consecutive run starting its walkers near the posterior maximum. The final run used 100 walkers per dimension each taking 25\,000 steps. Flat priors were used for each of $T_\mathrm{eff}$, $\log g$, and $\log(\mathrm{Si/H})$. The parallax value, $\varpi$, obtained from the \textit{Gaia} Data Release 3 (DR3) catalogue \citep{gaia2023dr3}, was used to obtain the distance to the source which is marginalised over and included as a parameter in the photometric fitting. The distance prior is a generalised gamma distribution following the Bailer-Jones method tailored to a white dwarf population \citep{bailer2021estimating}.

This provides us with measurements of $T_\mathrm{eff}$ and $\log g$ from both methods and the best-fit metallicity from the spectroscopic analysis. An iterative approach is used to try to achieve convergence between the two sets of parameters. The metallicity is fixed to the spectroscopic value in the photometric fit as the data are not sensitive to metals. Once we have an estimate for all metal abundances~--~determined as discussed in \hyperref[sec:abundance]{Section\,\ref*{sec:abundance}}~--~the 3D fitting can then be repeated with a new model grid, in which the metals vary by $\pm$ 1\,dex around the calculated abundances, and $T_\mathrm{eff}$ and $\log g$ can be recalculated.

\subsubsection{Spectroscopic Analysis}
For the spectroscopic analysis, we fitted the 3D model grid described in \hyperref[sec:models]{Section\,\ref*{sec:models}} which varies in $T_\mathrm{eff}$, $\log g$, and $\log(\mathrm{Si/H})$. The models are convolved by the resolution of the instrument; in this case $R = \lambda/\Delta\lambda = 5400$ for the UVB arm and $R = 8900$ for the VIS arm of the spectrograph. 

The spectra are first normalised because the flux calibration of ground-based data is affected by the Earth's atmosphere~--~this causes flux losses on the slit, particularly in bad seeing, coupled with differential refraction~--~and so the raw spectrum measured may not correspond to the physical distribution of intensity. For spectra with fewer absorption lines, the continuum usually serves as a reference point due to its invariance. Normalising sets it to unity which allows spectral lines to be measured in a consistent way. However, in the case of WD\,J0358 there is a high amount of metal absorption in the ultraviolet causing line blanketing \citep{mihalas1978stellar} and so it is difficult to define a continuum. We therefore adopt a method of normalisation used by e.g. \cite{dufour2012detailed}, in which the spectrum is normalised in windows of a fixed width (in this analysis we use 100\,\AA) using
\begin{equation}
    F_{\lambda_\mathrm{norm}} = \frac{F_\lambda}{\bar{F_\lambda}}\label{eq:1},
\end{equation}
where $F_{\lambda_\mathrm{norm}}$ is the normalised flux, $F_\lambda$ is the original flux, and $\bar{F_\lambda}$ is given by
\begin{equation}
    \bar{F_\lambda} = \sum_{i=1}^n \frac{F_{\lambda_i}}{n},
\end{equation}
with $n$ being the number of flux points. This is done for the full wavelength range of the spectrum and model. 

\begin{figure}
    \includegraphics[width=\columnwidth]{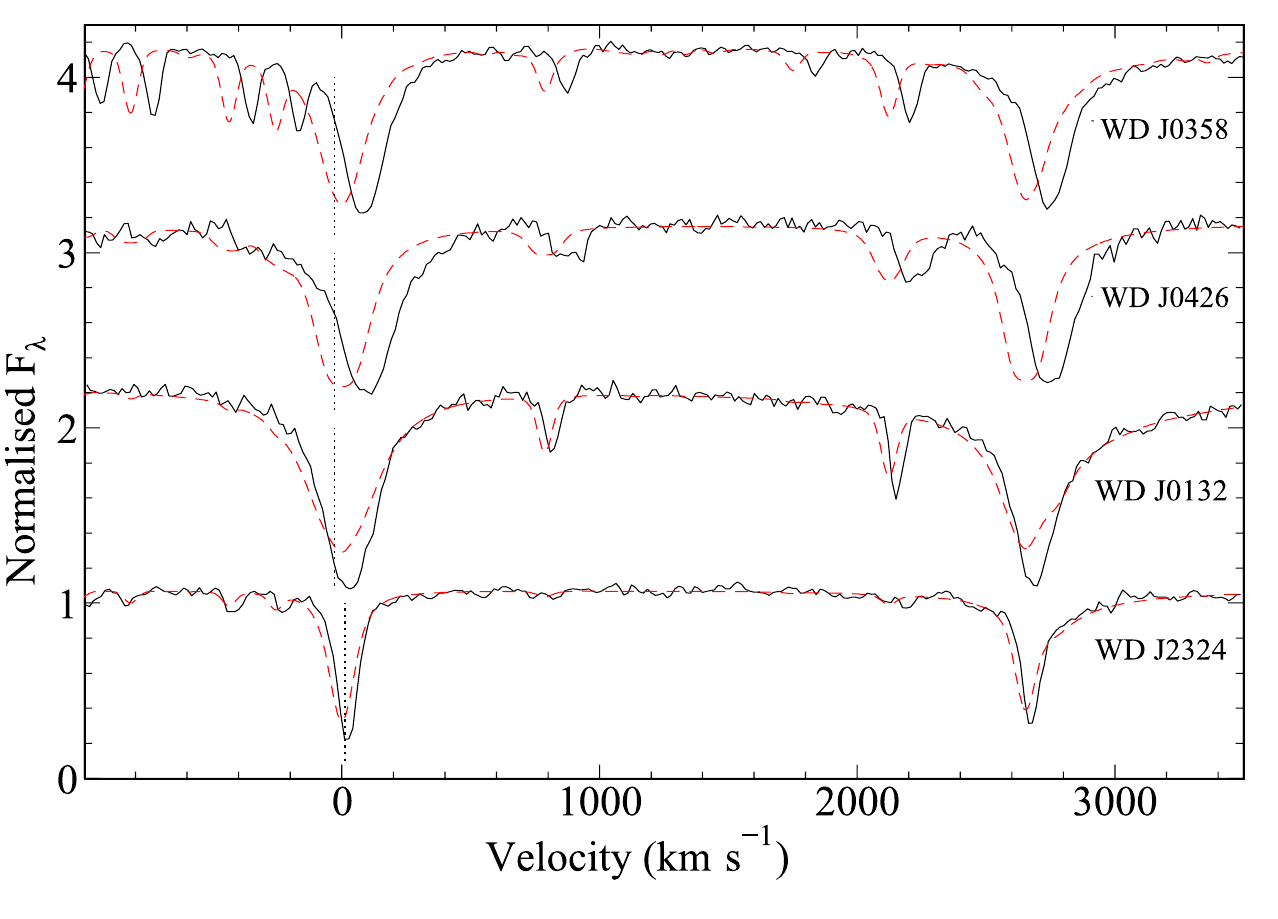}
    \caption{A section of \hyperref[fig:xshooter_spectra]{Fig.\,\ref*{fig:xshooter_spectra}} showing the X-Shooter data (black) and the models (red dashed) in the rest-frame of \ion{Ca}{ii}~3933.66\,\AA. The expected velocities of ISM absorption along the corresponding lines-of-sight are indicated by vertical dotted lines \citep{redfield2008structure}. For all stars, all detected lines share the same velocity offset with respect to the observed spectrum, indicating that there is no significant contribution from the ISM. The poor match of the Ca~H~\&~K lines in WD\,J0132 may indicate the presence of circumstellar material (see main text).}
    \label{fig:rv}
\end{figure}

We also applied a radial velocity (RV) correction to the models measured from the cross-correlation of the X-shooter spectrum and the model using a window that contains the strong Ca H~\&~K lines in the UVB data and the \ion{Ca}{II} 8600\,\AA\ triplet in the VIS data. We measured the RV shift for the UVB and VIS spectra, which are listed in \autoref{tab:wd_params}. A shift of $\simeq3.5\mathrm{km~s^{-1}}$ between the UVB and VIS arm is known \citep{sana2024xshooter}, the larger offset between the two velocities in WD\,J0426 is likely related to the magnetic nature of this star, and the offset seen in WD\,J0132 remains unexplained but is likely a data reduction issue. A comparison of the velocities of the Ca H~\&~K lines with those predicted for absorption in the local interstellar medium (ISM, see \hyperref[fig:rv]{Fig.\,\ref*{fig:rv}}) confirms that all detected metal lines are of photospheric origin (with the possible exception of WD\,J0132, which may show some contribution from circumstellar material). 

The MCMC routine is used to determine $T_\mathrm{eff}$, $\log g$, and $\log(\mathrm{Si/H})$ values by fitting the grid of synthetic model spectra to the X-shooter UVB and VIS spectra. The best-fit model parameter values are listed in \autoref{tab:wd_params}, along with their statistical uncertainties.

There is a puzzling discrepancy between the Ca H~\&~K lines in the best-fit model and the data for WD\,J0132. As can be seen in \hyperref[fig:rv]{Fig.\,\ref*{fig:rv}} and \hyperref[fig:metals_wdj0132]{Fig.\,\ref*{fig:metals_wdj0132}}, the observed absorption is deeper than predicted by the model, yet increasing the abundance of Ca causes the model lines to become broader, worsening the fit. We do not believe this is a resolution issue, as lines for other metals (e.g. Mg, Si, Fe) can be successfully fitted. There may be additional absorption along the line of sight, although it would be unusual to find ISM lines at these velocities \citep{redfield2008structure}. Circumstellar gas is another possibility, which has been detected in a small number of polluted white dwarfs (e.g. \citealt{debes2012detection, gansicke2012chemical, wilson2019multiwavelength, le2024revisiting, zuckerman2026high}). An additional unresolved absorption component may contribute to the unusually large ($\simeq15\,\mathrm{km~s^{-1}}$) offset in the velocities measured from the UVB and VIS arm (\autoref{tab:wd_params}).

\subsubsection{Photometric Analysis}
\label{sec:phot_fits}
For the photometric analysis, we used a 2D model grid with varying $T_\mathrm{eff}$ and $\log g$ as described in \hyperref[sec:models]{Section\,\ref*{sec:models}}, and fixed $\log(\mathrm{Si/H})$ to the spectroscopic value, since photometry cannot meaningfully constrain metal abundances. These model spectra are scaled by the solid angle of the star $4\pi(R/D)^2$, where $R$ is the radius of the white dwarf and $D$ is the distance to the white dwarf. We obtain $R$ from the white dwarf mass-radius relation using the \cite{bedard2020spectral} evolutionary models, assuming a thick H layer for the given $T_\mathrm{eff}$ and $\log g$ of each model, and $D$ by inverting the \textit{Gaia} DR3 parallax; this is then marginalised over in the fitting procedure. Existing photometric data were obtained from cross-matching the white dwarfs with the photometric surveys listed in \hyperref[sec:data]{Section\,\ref*{sec:data}} (Gentile Fusillo, private communication), and these were converted to fluxes. The photometric data used in the fitting are displayed in the appendix (\autoref{tab:phot_table}); however, we exclude the \textit{GALEX} photometry from the fitting as the fluxes will be affected by metal blanketing, and \textit{WISE} data are insufficient in spatial resolution and sensitivity~--~and potentially unreliable \citep{dennihy2020word}~--~for this fitting. Synthetic fluxes were calculated from the photometric bandpass of each of the photometric data points, obtained from the Spanish Virtual Observatory (SVO) Filter Profile Service \citep{rodrigo2012svo}. Due to the close distance of these targets, reddening is negligible, as determined from the 3D dust maps of \citet{green20193d}, and therefore is not included in our calculations.

\begin{figure}
    \includegraphics[width=\columnwidth]{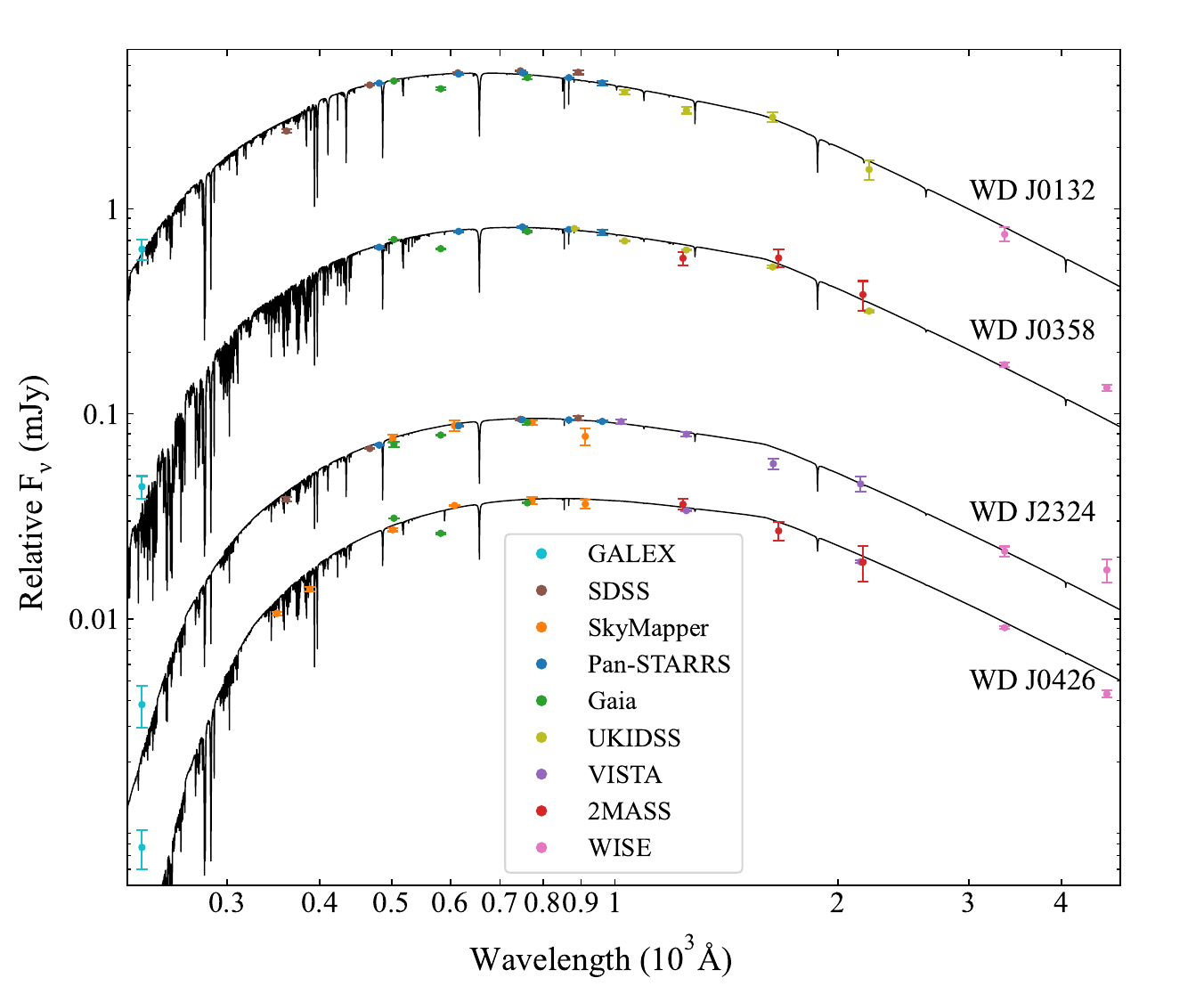}
    \caption{Spectral energy distributions (SEDs) for the four DAZs in order of temperature plotted with the corresponding photometric data. \textit{GALEX} and \textit{WISE} data are excluded from the photometric fitting. The spectra are vertically offset from WD\,J0358 for clarity.}
    \label{fig:sed}
\end{figure}

\subsubsection{Combined Fitting}
Comparing the results obtained from the spectroscopic and photometric fitting methods, we can see that $T_\mathrm{eff}$ differs outside of the error range for WD\,J0358 and WD\,J2324, whilst the other two agree with each other, and $\log g$ is only consistent between the two methods for WD\,J2324.

Due to these differences in the values of $T_\mathrm{eff}$ and $\log g$, we chose to test the hybrid fitting approach outlined in \cite{izquierdo2021gd} where the spectra and photometry are fitted simultaneously using the MCMC method. The best parameter values for this combined fitting are listed in \autoref{tab:wd_params}, along with their statistical uncertainties. We can see that in this method, $\log g$ has been constrained using the distance to the source and so is therefore similar to the purely photometric fit. The distance values also agree with the photometric value, as do the values of $\log(\mathrm{Si/H})$ to the spectroscopic value as expected (except for a small change in WD\,J2324). The hybrid values for $T_\mathrm{eff}$ do not follow any trend when compared to the values from the other methods.

In the subsequent analysis, we choose to use $T_\mathrm{eff}$, $\log g$, and distance calculated from the photometric MCMC fitting to construct our model atmospheres, as these are more reliable than those determined from either the spectroscopic or the hybrid technique (see e.g. Izquierdo et al., 2026). These $T_\mathrm{eff}$ and $\log g$ values are then used to infer the white dwarf mass ($M_\mathrm{WD}$) and cooling age ($\tau_\mathrm{cool}$) using white dwarf cooling models \citep{bedard2020spectral}; these are also listed in \autoref{tab:wd_params}. The errors quoted for each parameter determination method are purely the statistical output from the MCMC and do not include systematics, which are typically substantially larger \citep{izquierdo2023systematic}.

\begin{table*}
	\centering
	\caption{Atmospheric parameters determined from each fitting method. The 1D metallicity is defined in terms of $\log(\mathrm{Si/H})$ with the rest of the metals included at bulk Earth abundances relative to Si. For the photometric fitting, this is fixed to the value determined from the spectroscopic fit. We choose to use the photometric values in the subsequent analysis, and so we also include the white dwarf's mass, cooling age, and convection zone mass as calculated using photometric parameters.}
	\label{tab:wd_params}
	\begin{tabular}{lllllllllll}
		\hline
		\multicolumn{1}{c}{Star} & \multicolumn{1}{c}{Method} & \multicolumn{1}{c}{$T_\mathrm{eff}$} & \multicolumn{1}{c}{$\log g$} & \multicolumn{1}{c}{1D Metallicity} & \multicolumn{1}{c}{Distance} & \multicolumn{1}{c}{$M_\mathrm{WD}$} & \multicolumn{1}{c}{$\tau_\mathrm{cool}$} & \multicolumn{1}{c}{$M_\mathrm{CVZ}$} & $v_\mathrm{r,UVB}$ & $v_\mathrm{r,VIS}$ \\
            & & \multicolumn{1}{c}{(K)} & \multicolumn{1}{c}{(dex)} & \multicolumn{1}{c}{($\log(\mathrm{Si/H})$, dex)} & \multicolumn{1}{c}{(pc)} & \multicolumn{1}{c}{($M_\odot$)} & \multicolumn{1}{c}{(Gyr)} & \multicolumn{1}{c}{(g)} & \multicolumn{1}{c}{km\,s$^{-1}$} & \multicolumn{1}{c}{km\,s$^{-1}$}\\
		\hline
            \multirow{3}{*}{WD\,J0358} & Spec & $6554\pm12$ & $7.93\pm0.02$ & $-6.34\pm0.01$ & \textemdash\\
            & Phot & $6712\pm63$ & $8.19\pm0.02$ & $\mathit{-6.34}$ & $36.1\pm0.1$ & 0.70 & 2.45 & 7.66$\times10^{24}$ & 87.5 & 84.0 \\
            & Hybrid & $6687\pm10$ & $8.17\pm0.01$ & $-6.35\pm0.01$ & $36.4\pm0.1$\\
            \hline
            \multirow{3}{*}{WD\,J0426} & Spec & $6115\pm7$ & $8.39\pm0.01$ & $-7.53\pm0.01$ & \textemdash\\
            & Phot & $6032\pm18$ & $8.07\pm0.01$ & $\mathit{-7.53}$ & $34.4\pm0.1$ & 0.62 & 2.64 & 3.10$\times10^{25}$ & 94.4 & 82.7 \\
            & Hybrid & $6031\pm1$ & $8.06\pm0.01$ & $-7.52\pm0.01$ & $34.5\pm0.1$\\
            \hline
            \multirow{3}{*}{WD\,J0132} & Spec & $7312\pm18$ & $7.81\pm0.03$ & $-6.29\pm0.01$ & \textemdash\\
            & Phot & $7308\pm54$ & $7.93\pm0.03$ & $\mathit{-6.29}$ & $102.1\pm1.5$ & 0.55 & 1.27 & 9.25$\times10^{24}$ & 43.0 & 21.5 \\
            & Hybrid & $7338\pm14$ & $7.93\pm0.02$ & $-6.28\pm0.02$ & $102.3\pm1.2$\\
            \hline
            \multirow{3}{*}{WD\,J2324} & Spec & $6255\pm18$ & $7.93\pm0.02$ & $-7.81\pm0.02$ & \textemdash\\
            & Phot & $6421\pm51$ & $7.95\pm0.03$ & $\mathit{-7.81}$ & $82.1\pm1.2$ & 0.56 & 1.84 & 2.65$\times10^{25}$ & 21.5 & 26.4 \\
            & Hybrid & $6301\pm15$ & $7.93\pm0.01$ & $-7.75\pm0.03$ & $81.3\pm0.8$\\
		\hline
	\end{tabular}
\end{table*}

\subsection{Abundance Determination}
\subsubsection{Abundance Analysis}
\label{sec:abundance}
To measure metal abundances, once $T_\mathrm{eff}$ and $\log g$ were determined we fixed these parameters and computed 1D model grids for each element. We vary $\log(\mathrm{Z/H})$ by $\pm1\,$dex around the bulk Earth value in steps of $0.25$\,dex, while keeping all other metals fixed at their bulk Earth abundances relative to $\mathrm{Z}$, where $\mathrm{Z}$ is the selected metal for which we wish to obtain an abundance.

The fitting procedure utilises the line lists for the metals used in the model grid and fits a small 10\,\AA\ window centred on the strongest lines of each metal (listed in \autoref{tab:strong_lines}) in the spectra to find the abundance that minimises $\chi^2$. We chose to use minimum $\chi^2$ estimation for this fit as an MCMC routine would be excessive for this complexity of fit. The statistical errors on abundance measurements come from the interpolated abundance value at $\chi^2_\mathrm{min}+1$. The upper limits are calculated in a similar manner; we follow the method of \cite{hollands2020ultra} and convert the $\chi^2$ values to likelihoods using
\begin{equation}
    \ln{L} = -\frac{1}{2}\chi^2,
\end{equation}
then multiply this by the Jeffrey's prior, given by
\begin{equation}
    P(\log(\mathrm{Z/H})) = 10^{\log(\mathrm{Z/H})/2},
\end{equation}
and take the 99th percentile of the cumulative distribution function to be our upper limit value. The derived metal abundances and upper limits are listed in \autoref{tab:abundances}, along with the accretion rates for each, assuming convective overshoot \citep{cunningham2019convective} and steady state accretion. The individual metal accretion rates are represented by the diffusion fluxes of the metals sinking out of the convection zone. These are computed using the measured photospheric abundances combined with the models describing the diffusion of metals from the surface; the models account for gravitational settling, chemical diffusion, thermal diffusion, and radiative diffusion \citep{paquette1986diffusion, dupuis1992study, koester2009accretion}. The metals sink out of the convection zone at different rates, and we convert the photospheric abundances to parent body abundances as described in \hyperref[sec:accretion]{Section\,\ref*{sec:accretion}}, assuming steady state. 

\begin{table*}
    \centering
    \caption{Photospheric number abundances of metals found in the four DAZs and corresponding diffusion times, steady state accretion rates, and masses. O is the only major constituent of rocky material not detected in any of the white dwarfs and so an upper limit is included, as well as upper limits for other missing metals from the eleven detected in WD\,J0358.}
    \label{tab:abundances}
    \begin{tabular}{lcccccccc}
	\hline
        \multicolumn{1}{l}{Z} & \multicolumn{1}{c}{$\log(\mathrm{Z/H})$} & \multicolumn{1}{c}{$\tau_{\mathrm{Z}}$} & \multicolumn{1}{c}{$\dot{M}_{\mathrm{Z}}$} & \multicolumn{1}{c}{$M_{\mathrm{Z}}$} & \multicolumn{1}{c}{$\log(\mathrm{Z/H})$} & \multicolumn{1}{c}{$\tau_{\mathrm{Z}}$} & \multicolumn{1}{c}{$\dot{M}_{\mathrm{Z}}$} & \multicolumn{1}{c}{$M_{\mathrm{Z}}$}\\
        & \multicolumn{1}{c}{(dex)} & \multicolumn{1}{c}{(yr)} & \multicolumn{1}{c}{(g~s$^{-1}$)} & \multicolumn{1}{c}{(g)} & \multicolumn{1}{c}{(dex)} & \multicolumn{1}{c}{(yr)} & \multicolumn{1}{c}{(g~s$^{-1}$)} & \multicolumn{1}{c}{(g)}\\
        \hline
        & \multicolumn{4}{c}{\textbf{WD\,J0358}} & \multicolumn{4}{c}{\textbf{WD\,J0426}}\\
	\hline
        O & $<-3.59$ & \textemdash & \textemdash & \textemdash & $<-2.37$ & \textemdash & \textemdash & \textemdash\\
        Na & $-8.50\pm0.11$ & 5.32$\times10^3$ & 3.26$\times10^6$ & 1.73$\times10^{10}$ & $-8.34\pm0.06$ & 1.82$\times10^4$ & 5.65$\times10^6$ & 1.03$\times10^{11}$\\
        Mg & $-6.34\pm0.01$ & 5.26$\times10^3$ & 5.04$\times10^8$ & 2.65$\times10^{12}$ & $-7.20\pm0.01$ & 1.80$\times10^4$ & 8.31$\times10^7$ & 1.50$\times10^{12}$\\
        Al & $-7.76\pm0.03$ & 4.89$\times10^3$ & 2.28$\times10^7$ & 1.12$\times10^{11}$ & $-8.23\pm0.02$ & 1.68$\times10^4$ & 9.22$\times10^6$ & 1.55$\times10^{11}$\\
        Si & $-6.28\pm0.05$ & 4.88$\times10^3$ & 7.21$\times10^8$ & 3.51$\times10^{12}$ & $-7.15\pm0.09$ & 1.68$\times10^4$ & 1.15$\times10^8$ & 1.94$\times10^{12}$\\
        Ca & $-7.86\pm0.01$ & 3.94$\times10^3$ & 3.35$\times10^7$ & 1.32$\times10^{11}$ & $-8.51\pm0.01$ & 1.39$\times10^4$ & 8.74$\times10^6$ & 1.21$\times10^{11}$\\
        Ti & $-9.18\pm0.07$ & 3.39$\times10^3$ & 2.23$\times10^6$ & 7.54$\times10^9$ & $<-8.97$ & \textemdash & \textemdash & \textemdash\\
        Cr & $-7.98\pm0.03$ & 3.22$\times10^3$ & 4.03$\times10^7$ & 1.30$\times10^{11}$ & $-9.19\pm0.03$ & 1.14$\times10^4$ & 2.88$\times10^6$ & 3.28$\times10^{10}$\\
        Mn & $-8.71\pm0.05$ & 3.09$\times10^3$ & 8.27$\times10^6$ & 2.55$\times10^{10}$ & $-9.57\pm0.07$ & 1.10$\times10^4$ & 1.32$\times10^6$ & 1.45$\times10^{10}$\\
        Fe & $-6.17\pm0.01$ & 3.08$\times10^3$ & 2.92$\times10^9$ & 9.00$\times10^{12}$ & $-7.55\pm0.01$ & 1.10$\times10^4$ & 1.40$\times10^8$ & 1.54$\times10^{12}$\\
        Co & $-8.87\pm0.11$ & 2.95$\times10^3$ & 6.41$\times10^6$ & 1.89$\times10^{10}$ & $-9.42\pm0.07$ & 1.05$\times10^4$ & 2.08$\times10^6$ & 2.19$\times10^{10}$\\
        Ni & $-7.44\pm0.02$ & 3.01$\times10^3$ & 1.69$\times10^8$ & 5.08$\times10^{11}$ & $-8.78\pm0.02$ & 1.07$\times10^4$ & 8.87$\times10^6$ & 9.52$\times10^{10}$\\
        Total & \textemdash & \textemdash & 4.43$\times10^9$ & 1.61$\times10^{13}$ & \textemdash & \textemdash & 3.78$\times10^8$ & 5.52$\times10^{12}$\\
	\hline
        & \multicolumn{4}{c}{\textbf{WD\,J0132}} & \multicolumn{4}{c}{\textbf{WD\,J2324}}\\
        \hline
        O & $<-3.56$ & \textemdash & \textemdash & \textemdash & $<-2.88$ & \textemdash & \textemdash & \textemdash\\
        Na & $<-8.03$ & \textemdash & \textemdash & \textemdash & $<-8.18$ & \textemdash & \textemdash & \textemdash\\
        Mg & $-6.29\pm0.03$ & 1.11$\times10^4$ & 3.26$\times10^8$ & 3.63$\times10^{12}$ & $-8.01\pm0.08$ & 2.08$\times10^4$ & 9.49$\times10^6$ & 1.98$\times10^{11}$\\
        Al & $-7.18\pm0.07$ & 1.03$\times10^4$ & 5.03$\times10^7$ & 5.19$\times10^{11}$ & $-8.93\pm0.20$ & 1.94$\times10^4$ & 1.08$\times10^6$ & 2.64$\times10^{10}$\\
        Si & $-6.63\pm0.15$ & 1.03$\times10^4$ & 1.87$\times10^8$ & 1.92$\times10^{12}$ & $<-7.09$ & \textemdash & \textemdash & \textemdash\\
        Ca & $-6.97\pm0.01$ & 8.18$\times10^3$ & 1.53$\times10^8$ & 1.25$\times10^{12}$ & $-8.76\pm0.02$ & 1.58$\times10^4$ & 3.68$\times10^6$ & 5.80$\times10^{10}$\\
        Ti & $-8.47\pm0.15$ & 7.00$\times10^3$ & 6.75$\times10^6$ & 4.72$\times10^{10}$ & $<-9.45$ & \textemdash & \textemdash & \textemdash\\
        Cr & $-8.72\pm0.17$ & 6.63$\times10^3$ & 4.35$\times10^6$ & 2.88$\times10^{10}$ & $<-9.15$ & \textemdash & \textemdash & \textemdash\\
        Mn & $<-8.63$ & \textemdash & \textemdash & \textemdash & $<-9.38$ & \textemdash & \textemdash & \textemdash\\
        Fe & $-7.30\pm0.01$ & 6.33$\times10^3$ & 1.15$\times10^8$ & 7.26$\times10^{11}$ & $-7.53\pm0.01$ & 1.24$\times10^4$ & 1.11$\times10^8$ & 1.37$\times10^{12}$\\
        Co & $<-8.72$ & \textemdash & \textemdash & \textemdash & $<-9.39$ & \textemdash & \textemdash & \textemdash\\
        Ni & $<-8.62$ & \textemdash & \textemdash & \textemdash & $-8.95\pm0.08$ & 1.21$\times10^4$ & 4.53$\times10^6$ & 5.49$\times10^{10}$\\
        Total & \textemdash & \textemdash & 8.42$\times10^8$ & 8.12$\times10^{12}$ & \textemdash & \textemdash & 1.30$\times10^8$ & 1.71$\times10^{12}$\\
        \hline
	\end{tabular}
\end{table*}

\subsubsection{Magnetic Fitting}
\begin{figure*}
\centering
\begin{subfigure}{.6\textwidth}
  \centering
  \includegraphics[width=\linewidth]{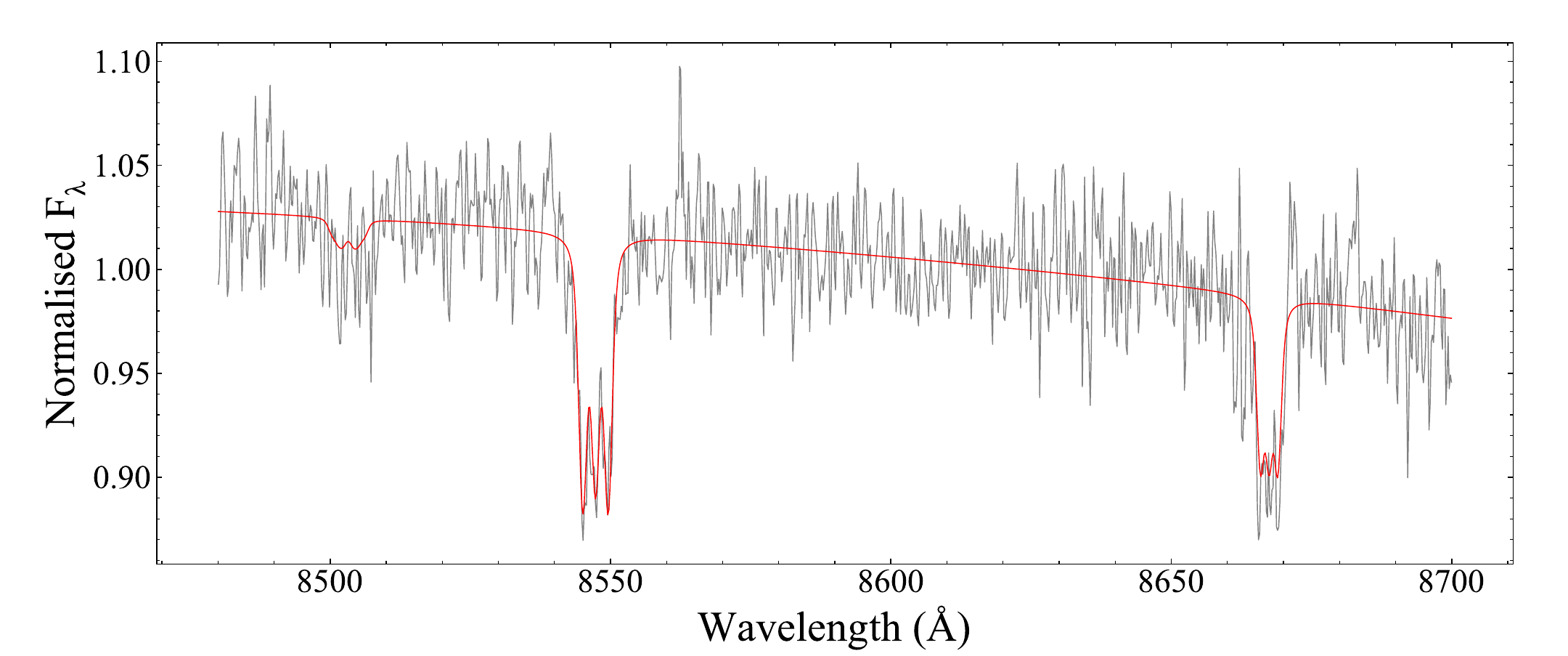}
  \label{fig:mag_sub1}
\end{subfigure}%
\hfill
\begin{subfigure}{.4\textwidth}
  \centering
  \includegraphics[width=\linewidth]{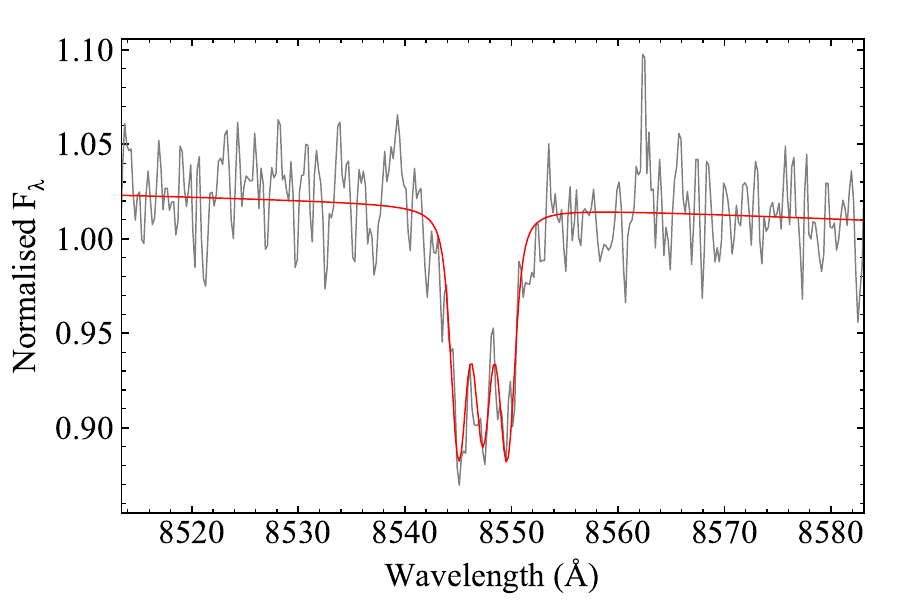}
  \label{fig:mag_sub2}
\end{subfigure}
\caption{Left: \ion{Ca}{ii} triplet in WD\,J0426 displaying Zeeman splitting caused by a 58~kG magnetic field. Right: A zoom-in on the 8542\,\AA\ line. The best-fit magnetic model is overlaid in red.}
\label{fig:mag_ca}
\end{figure*}

We detected Zeeman splitting in the metal lines of WD\,J0426 (see e.g. the \ion{Ca}{II} triplet in \hyperref[fig:mag_ca]{Fig.\,\ref*{fig:mag_ca}}), and hence we re-classify this star as a DAZH white dwarf (where H denotes magnetism), contrary to its previous classification as a DAZ \citep{obrien2023gaia}. This is the reddest multiplet, and thus the most affected by the magnetic field because the splitting scales with $\lambda^2$ \citep{gray2021observation}. Therefore, we calculate the strength of the magnetic field by fitting this region with synthetic spectra of the triplet\footnote{https://github.com/mahollands/magnetic/} varying magnetic field strength, Gaussian FWHM, Lorentzian FWHM, viewing angle, and RV using non-linear least squares estimation. We find the magnetic field to be $B=58\pm0.2\,\mathrm{kG}$.

Weak magnetic fields have been shown not to affect the thermal atmosphere structure of the white dwarf, and so the determination of atmospheric parameters without accounting for magnetism is still valid \citep{farihi2011magnetic, zuckerman2011aluminum}. However, the effect of Zeeman splitting must be included in synthetic spectra in order to obtain accurate metal abundance measurements. We therefore assume the linear Zeeman effect (which dominates in the low-field regime) in a constant B-field of $\simeq60\,\mathrm{kG}$, which we apply to all detected metal multiplets in order to create magnetic model spectra to fit to the data. 

An atomic level in this regime with total angular momentum $J$ splits into $2J+1$ components, each with magnetic quantum number $m_\mathrm{J}=-J, ..., +J$ in integer steps. Electric dipole transitions were calculated between the lower and upper states allowed by the selection rules $\Delta J=0, \pm1$ and $\Delta m_\mathrm{J}=0, \pm1$. The energies of these states are then shifted to
\begin{equation}
    k=k_0+46.686g_\mathrm{J}m_\mathrm{J}B,
\end{equation}
where $k_0$ is the rest energy in $\mathrm{cm}^{-1}$, 46.686 is the Bohr magneton in $\mathrm{cm}^{-1}~\mathrm{MG}^{-1}$, $g_\mathrm{J}$ is the Land\'e $g$-factor, $m_\mathrm{J}$ is the magnetic quantum number, and $B$ is the magnetic field strength in G. The Land\'e $g$-factor can be calculated from the three angular momentum quantum numbers $J$, $L$, and $S$ using \citep{lande1923termstruktur}
\begin{equation}
    g_\mathrm{J}=1+\frac{J(J+1)+S(S+1)-L(L+1)}{2J(J+1)}.
\end{equation}
The relative strengths of the $\pi$ and $\sigma$ ($\Delta m_\mathrm{J}=0, \pm1$, respectively) Zeeman components are calculated following \cite{honl1925intensitaten} (see also Table B.1. from \citealt{dorsch2022discovery}). The atomic data for all multiplets are obtained from the NIST Atomic Spectra Database \citep{NIST_ASD} and following this methodology, the wavelengths of the split components and their oscillator strengths are added to our element line lists to create magnetic model spectra. These are fitted in the same manner as \hyperref[sec:abundance]{Section\,\ref*{sec:abundance}} to obtain the abundance values and upper limits listed in \autoref{tab:abundances}. As shown in \hyperref[fig:mag_ca]{Fig.\,\ref*{fig:mag_ca}}, the magnetic model fitted well to the data, despite using a uniform B-field over the whole wavelength range to simplify the analysis.

\subsubsection{Diffusion and Accretion Calculations}
Another parameter of interest is the total accretion rate, $\dot{M}_\mathrm{Z_{tot}}$, which is simply the sum of the individual accretion rates for all detected metals and is therefore a minimum accretion rate, especially with no O detections which can increase the total by up to two orders of magnitude (e.g. \citealt{swan2019interpretation}). The (steady-state) accretion rate of each metal is calculated as in \cite{koester2009accretion} by
\begin{equation}
    \dot{M}_\mathrm{Z} = \frac{M_\mathrm{CVZ}X_\mathrm{Z}}{\tau_\mathrm{Z}}\label{eq:7},
\end{equation}
where $M_\mathrm{CVZ}$ is the mass in the convection zone, $X_\mathrm{Z}$ is the mass abundance of each metal, and $\tau_\mathrm{Z}$ is the diffusion time for each detected metal. This equation describes the diffusion flux out of the base of the convection zone, however, in the steady state the accretion and diffusion rates balance. The total (and therefore minimum) accreted mass of the material, $M_\mathrm{Z_{min}}$, is given by
\begin{equation}
    M_\mathrm{Z_{min}} = \sum_\mathrm{Z}\tau_\mathrm{Z}\dot{M}_\mathrm{Z}.
\end{equation}
The total accretion rate and minimum accreted mass for each DAZ are provided in \autoref{tab:abundances}.

We subsequently use our calculated abundances and corresponding accretion rates/masses of the detected metals to characterise the nature of the accreted material in each of the four DAZs. We compare our photospheric metal compositions to the compositions of bodies in our Solar System; this includes the abundances of individual meteorites \citep{nittler2004bulk}, Earth \citep{wang2018elemental}, Mars \citep{yoshizaki2020composition}, CI chondrites \citep{lodders2019solar}, and the Sun \citep{lodders2019solar}.

\subsubsection{Accretion Phases}
\label{sec:accretion}
In this analysis, we use the model for accretion history described in \cite{koester2009accretion}, in which accretion switches on, remains at a constant rate until all the material has been transferred onto the white dwarf, and then switches off. This causes there to be three different phases of accretion over the event lifetime: an increasing phase, steady state, and a decreasing phase. In this scenario, the measured metal abundances may not correspond directly to the parent body abundances, depending on which phase of accretion the planetary material is in.

At the start of the increasing phase, the planetary material is being accreted onto the white dwarf but not enough time has passed for the metals to diffuse out of the convection zone. This means that the measured abundances will directly mirror the composition of the parent body until the elements begin to settle, i.e., 
\begin{equation}
    \left(\frac{X(Z_1)}{X(Z_2)}\right)_\mathrm{par} = \left(\frac{X(Z_1)}{X(Z_2)}\right)_\mathrm{WD},
\end{equation}
where $X(Z_1)/X(Z_2)$ is an abundance ratio for two elements in the parent body, par, and white dwarf, WD.
In steady state, the metals diffuse out of the convection zone at the rate of their respective sinking timescales, and this is in equilibrium with the rate at which they are being accreted, and so the metal abundance remains constant. Heavier elements, e.g. Fe, settle quicker than lighter elements, e.g. Na, \citep{paquette1986diffusion} so the measured abundances may appear depleted in siderophiles and over-abundant in lithophiles. We correct for this using the sinking times for each element in the ratio, $\tau_\mathrm{Z_1}$ and $\tau_\mathrm{Z_2}$, i.e.,
\begin{equation}
    \left(\frac{X(Z_1)}{X(Z_2)}\right)_\mathrm{par} = \left(\frac{X(Z_1)}{X(Z_2)}\right)_\mathrm{WD} \left(\frac{\tau_\mathrm{Z_2}}{\tau_\mathrm{Z_1}}\right)\label{eq:10}.
\end{equation}
In the decreasing phase, accretion has ceased and all the material continues to diffuse out of the convection zone exponentially at their respective rates, causing a further divergence in the detected abundances of heavier and lighter elements. This relates to the parent body using
\begin{equation}
    \left(\frac{X(Z_1)}{X(Z_2)}\right)_\mathrm{par} = \left(\frac{X(Z_1)}{X(Z_2)}\right)_\mathrm{WD} \left(\frac{e^{-t/\tau_\mathrm{Z_2}}}{e^{-t/\tau_\mathrm{Z_1}}}\right),
\end{equation}
where $t$ is the time since accretion ended.

It is difficult to determine what accretion phase the white dwarf is in, especially for He atmospheres as the metals can remain visible in the spectra long after accretion has ended. For H atmospheres, it is generally assumed the white dwarf is in the steady state with ongoing accretion as the metals sink out of the photosphere on much faster timescales~--~this is what we assume for the subsequent analysis.

\section{Results}
\subsection{Photospheric Metals}
We analysed the four DAZs following the methods outlined in \hyperref[sec:analysis]{Section\,\ref*{sec:analysis}} and found them to contain multiple elements in their photospheres, the least being five. As such, they fall into the minority of polluted white dwarfs in which more than one element is detected (see Fig.\,5. in \citealt{williams2024pewdd}).

\begin{figure*}
    \includegraphics[width=\textwidth]{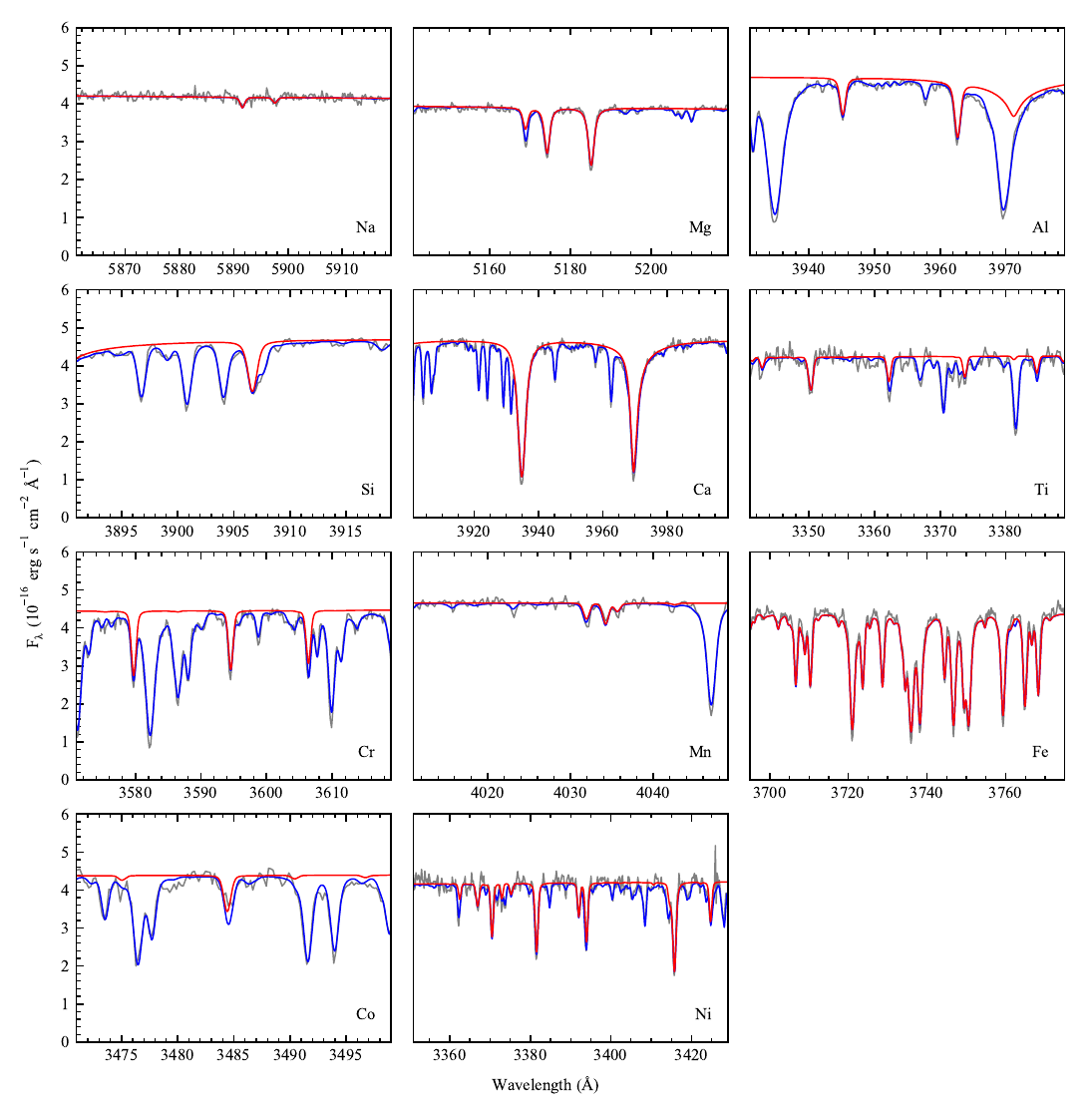}
    \caption{Absorption lines for all metals detected in WD\,J0358. The best-fit model spectrum is plotted over the data in blue and the individual lines for each metal are denoted by the red model.}
    \label{fig:metals_wdj0358}
\end{figure*}

\begin{figure*}
    \includegraphics[width=\textwidth]{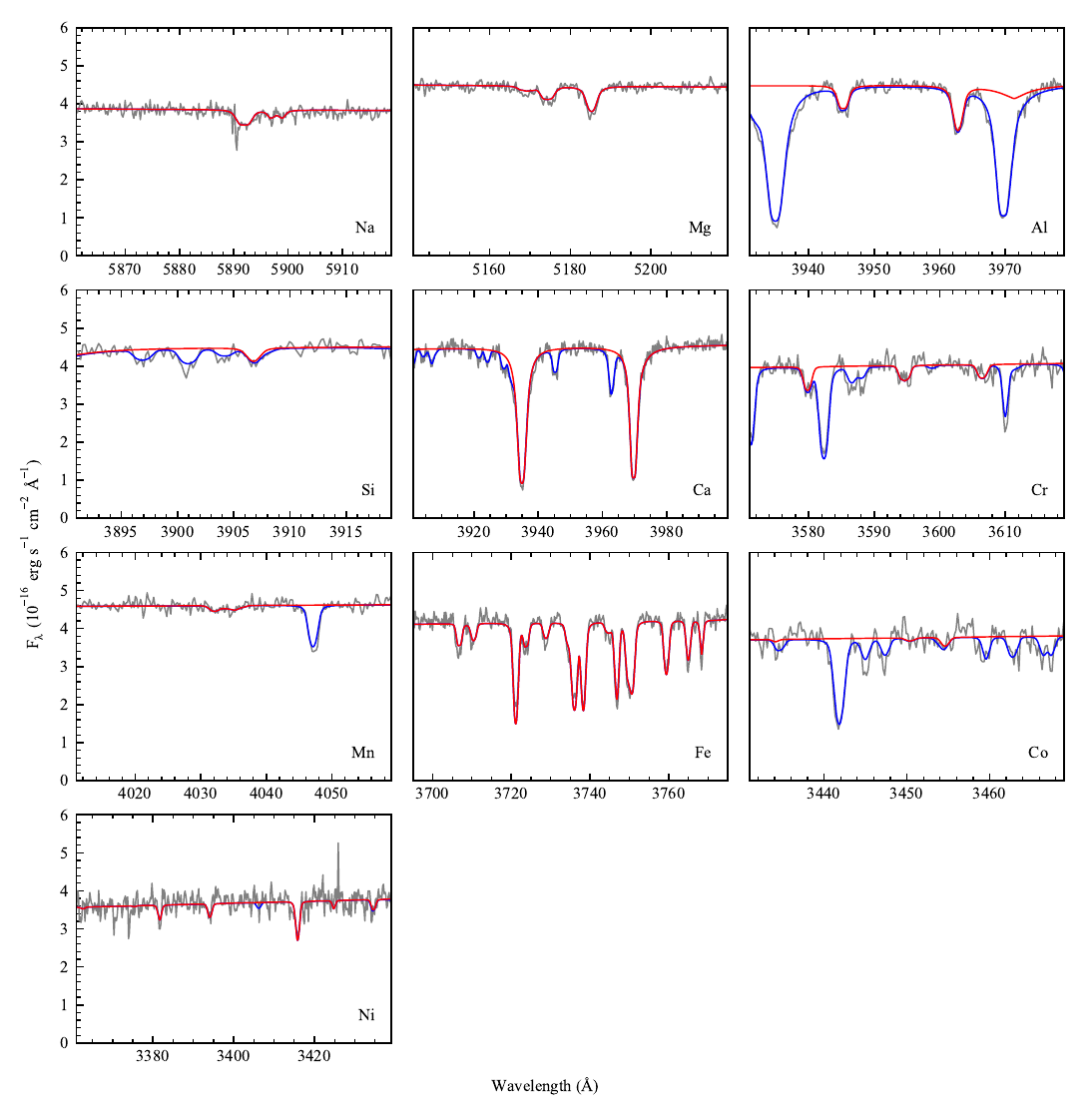}
    \caption{Absorption lines for all metals detected in WD\,J0426. The best-fit model spectrum is plotted over the data in blue and the individual lines for each metal are denoted by the red model.}
    \label{fig:metals_wdj0426}
\end{figure*}

\begin{figure*}
    \includegraphics[trim={0 4.5cm 0 0}, clip, width=\textwidth]{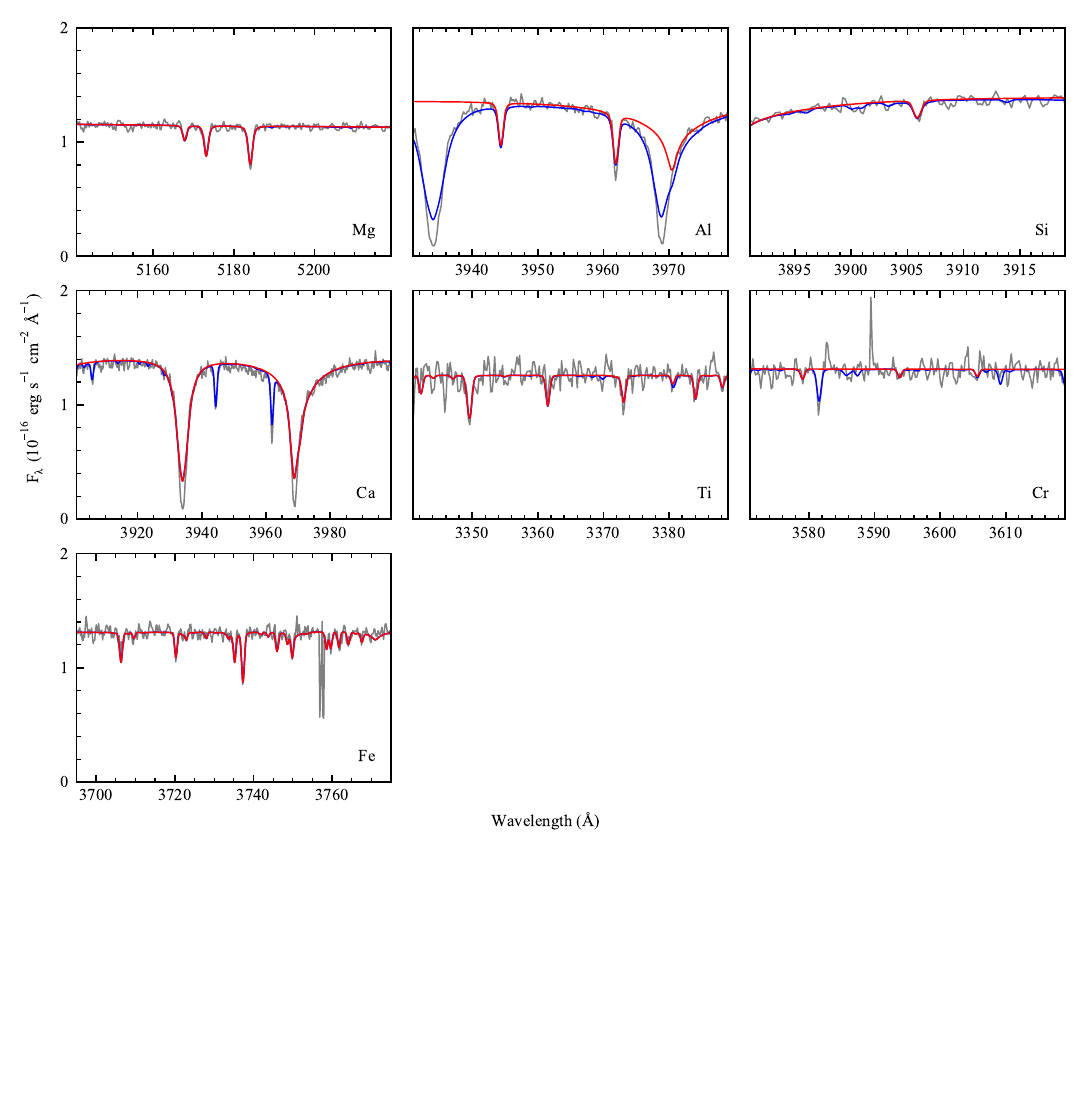}
    \caption{Absorption lines for all metals detected in WD\,J0132. The best-fit model spectrum is plotted over the data in blue and the individual lines for each metal are denoted by the red model.}
    \label{fig:metals_wdj0132}
\end{figure*}

\begin{figure*}
    \includegraphics[trim={0 8.5cm 0 0}, clip, width=\textwidth]{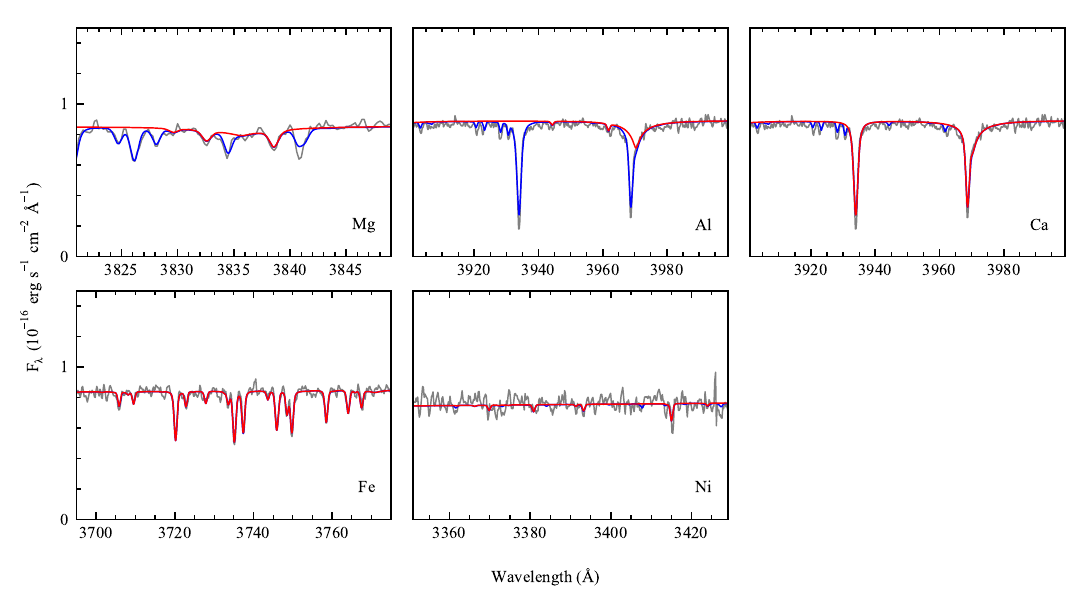}
    \caption{Absorption lines for all metals detected in WD\,J2324. The best-fit model spectrum is plotted over the data in blue and the individual lines for each metal are denoted by the red model.}
    \label{fig:metals_wdj2324}
\end{figure*}

The WD\,J0358 spectra show a high amount of metal pollution; we identify eleven metals and give an upper limit for O as the only missing major bulk Earth metal, because O lines are undetectable in the optical at these temperatures. Strong lines for each of the eleven metals are shown in \hyperref[fig:metals_wdj0358]{Fig.\,\ref*{fig:metals_wdj0358}}. For the other three white dwarfs, we also give upper limits for any of these elements with no detection. We detect ten metals in WD\,J0426, seven metals in WD\,J0132, and five metals in WD\,J2324, shown in \hyperref[fig:metals_wdj0426]{Fig.\,\ref*{fig:metals_wdj0426}}, \hyperref[fig:metals_wdj0132]{Fig.\,\ref*{fig:metals_wdj0132}}, and \hyperref[fig:metals_wdj2324]{Fig.\,\ref*{fig:metals_wdj2324}}, respectively. These values are listed in \autoref{tab:abundances}.


\subsection{Abundance Ratios}
\label{sec:abundance_ratios}
Using the metal abundances from \autoref{tab:abundances} and correcting for differential sinking times in the steady state using \hyperref[eq:10]{Eq.\,\ref*{eq:10}}, we produce abundance ratio plots for our four DAZs along with Solar System objects to give a visual assessment of how similar the accreted material compositions are to these known bodies.

\begin{figure}
    \includegraphics[width=\columnwidth]{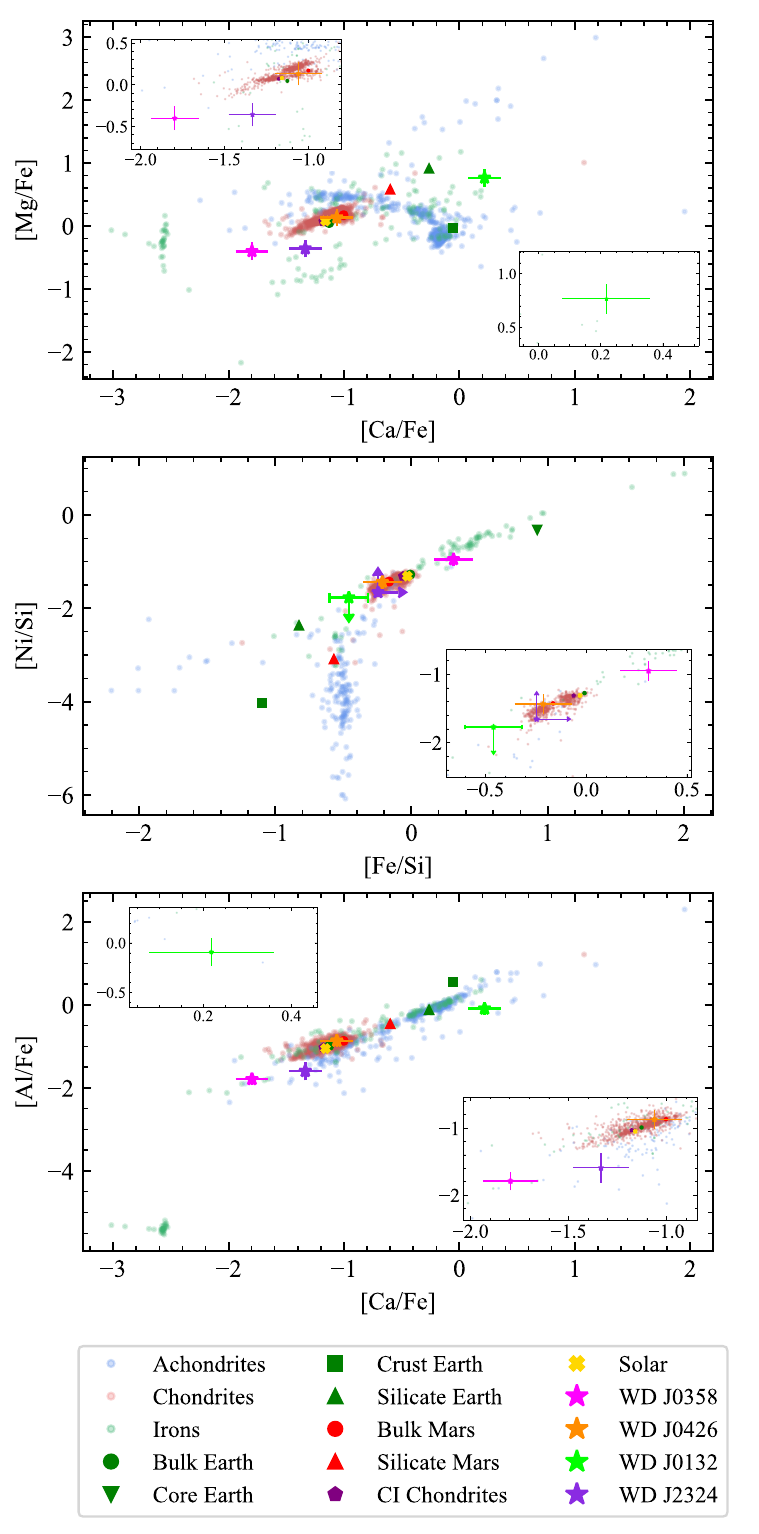}
    \caption{Abundance ratios plotting core, mantle, and crust tracers in the top panel; siderophiles in the middle panel; and lithophiles in the bottom panel. Each plot contains Solar System bodies alongside the four DAZs for comparison. Zoom-ins are provided to display the DAZ abundance errors; we use a 0.1 error in the propagation for any abundance with an error less than this, in line with the literature.}
    \label{fig:abundance_ratios}
\end{figure}

The top panel of \hyperref[fig:abundance_ratios]{Fig.\,\ref*{fig:abundance_ratios}} shows the elements that are the main tracers of core, mantle, and crust (Fe, Mg, and Ca, respectively). WD\,J0426 lies in the same region as the bulk planets and CI chondrites, whereas both WD\,J0358 and WD\,J2324 seem to be depleted in crust/mantle-like material. On the opposite end, WD\,J0132 follows the trend of primitive mantles and continental crust with a higher amount of Ca detected, implying a more crust/mantle-like composition of its parent body.

The middle panel plots siderophiles Ni and Fe relative to Si. These elements are expected to be found in the core of a differentiated planetesimal, so we can deduce from the plot that WD\,J0358 has accreted very core-rich material from its high Ni and Fe content. Both WD\,J0426 and WD\,J2324 lie in the bulk material region, however Si is only an upper limit for the latter and so this does not constrain it's composition. WD\,J0132 is lacking in core elements; it contains less Fe than the bulk planets and CI chondrites, and Ni is only an upper limit here.

The bottom panel shows lithophiles Al and Ca relative to Fe. These elements are indicative of material from the crust of a differentiated body, and we can see that WD\,J0132 does indeed have crust-like abundances, although with a greater abundance of Ca than the Earth's crust. As previously, WD\,J0426 and WD\,J2324 match the abundances of the bulk Solar System bodies, but WD\,J0358 has a depletion of these elements in comparison.

\subsection{Oxygen Upper Limits}
\autoref{tab:abundances} contains upper limits for O for the four DAZs, calculated as described in \hyperref[sec:abundance]{Section\,\ref*{sec:abundance}}. O is a major constituent of the bulk Earth composition, and it is known that in the Solar System, metals in rocky bodies primarily come in the form of oxides, so it is important to consider when deriving the mass fractions of metals. However, our weak upper limits provide no meaningful constraint on the O content of the accreted material. We therefore follow the method introduced in \cite{klein2010chemical} to calculate the amount of O needed to form oxides $Z_{p}\mathrm{O}_{q}$, given by
\begin{equation}
    \frac{n(\mathrm{O})}{n(\mathrm{H})} = \sum_{Z}\frac{q}{p}\frac{n(Z)}{n(\mathrm{H})},
    \label{eq:12}
\end{equation}
where $p$ is the number of atoms of element $Z$ and $q$ is the number of O atoms in the oxide.

We account for the oxides \ce{Na2O}, \ce{MgO}, \ce{Al2O3}, \ce{SiO2}, \ce{CaO}, \ce{TiO2}, \ce{Cr2O3}, \ce{MnO}, \ce{FeO}, \ce{CoO}, and \ce{NiO}; Mg, Si and Fe oxides will dominate since they constitute the majority of the material. Fe can potentially come in three different states: metallic Fe, \ce{FeO}, and the more O-rich alloy, \ce{Fe2O3}. Since the oxidation state of Fe in the accreted parent body is unknown and none of our stars have an O detection, we exclude the O-rich form of Fe and make a simplifying assumption that \ce{FeO} is the only oxide that can form (see \citealt{williams2025measurements}), as well as considering Fe in its metallic form. 

\autoref{tab:oxygen} gives the expected O abundances in each of the stars in the cases of Fe all in the form \ce{FeO}, all in metallic form, and half \ce{FeO}-half Fe. We only include detected metals in our calculations, with the exception of WD\,J2324, which is also missing another major constituent: Si. In this case, we estimate a $\log(\mathrm{Si/H})$ assuming a ratio of $\log(\mathrm{Si/Mg})=-0.04$ in line with bulk Earth abundances. We use the O value assuming half of Fe is in the form of \ce{FeO}, and half in its metallic form (Fe) in our subsequent analysis. These values assume all the accreted body was dry rock and does not account for the possibility of water ice being accreted which would increase the amount of O present.

\begin{table}
    \centering
    \caption{O abundances calculated for different forms of Fe, assuming that it was delivered in metal oxides in dry rock}.
    \label{tab:oxygen}
    \begin{tabular}{lcccc}
	\hline
        \multicolumn{1}{c}{Form of Fe} & \multicolumn{4}{c}{$\log(\mathrm{O/H})$}\\
        & \multicolumn{1}{c}{\textbf{WD\,J0358}} & \multicolumn{1}{c}{\textbf{WD\,J0426}} & \multicolumn{1}{c}{\textbf{WD\,J0132}} & \multicolumn{1}{c}{\textbf{WD\,J2324}}\\
        \hline
         All \ce{FeO} & $-$5.64 & $-$6.60 & $-$5.91 & $-$7.21\\
         All metallic Fe & $-$5.79 & $-$6.65 & $-$5.92 & $-$7.49\\
         $\frac{1}{2}$\ce{FeO} $\frac{1}{2}$Fe & $-$5.71 & $-$6.63 & $-$5.91 & $-$7.33\\
        \hline
    \end{tabular}
\end{table}

\subsection{Mass Fractions}
\label{sec:mass_fractions}
\begin{figure*}
    \includegraphics[width=\textwidth]{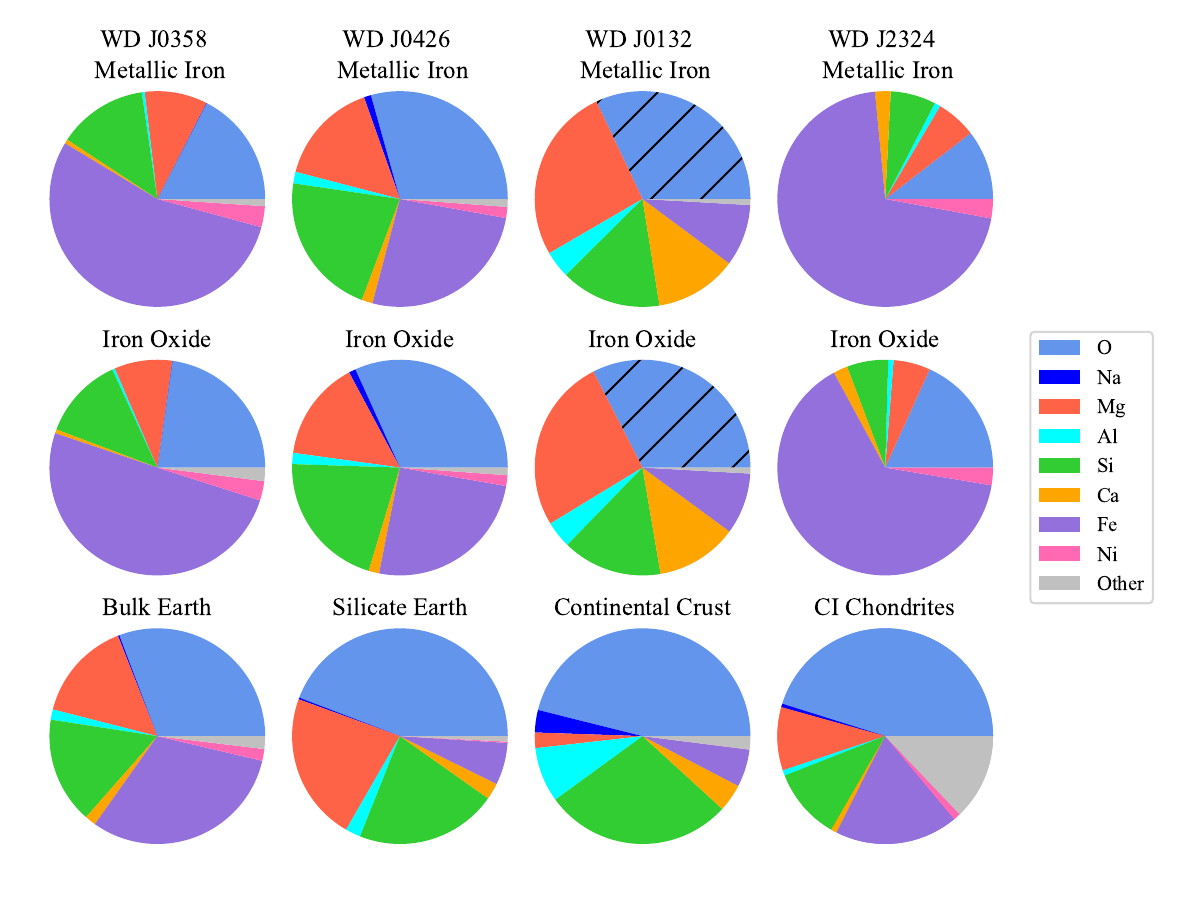}
    \caption{Mass fractions for the four DAZs, assuming O was present in the parent body in metal oxides, with the top row including only metallic Fe and middle row only \ce{FeO}. The bottom row shows bulk Earth, silicate Earth, continental crust, and CI chondrite mass fractions for comparison. The hashed portions indicate an estimated mass fraction for any missing major constituents of rocky material.}
    \label{fig:mass_fractions}
\end{figure*}

With estimates for O in the four systems, we can calculate mass fractions from the compositions. We display these in \hyperref[fig:mass_fractions]{Fig.\,\ref*{fig:mass_fractions}}, showing the major constituents of rocky material, alongside any other metals that were present in each, with estimated mass fractions shown by hashed sections. For comparison, we also show the mass fractions for bulk Earth, silicate Earth, continental crust, and CI chondrites. Mass fractions for the metals when Fe is in the form half \ce{FeO}-half Fe are used in our subsequent analysis and listed in \autoref{tab:mass_frac}.

\begin{table}
    \centering
    \caption{Mass fractions of metals using the O abundance calculated using \hyperref[eq:12]{Eq.\,\ref*{eq:12}} with half of Fe in the form \ce{FeO} and half as Fe. Estimates for metals with no photospheric detection are indicated by italics.}
    \label{tab:mass_frac}
    \begin{tabular}{lcccc}
	\hline
        \multicolumn{1}{l}{Z} & \multicolumn{4}{c}{Mass Fraction (per cent)}\\
        & \multicolumn{1}{c}{WD\,J0358} & \multicolumn{1}{c}{WD\,J0426} & \multicolumn{1}{c}{WD\,J0132} & \multicolumn{1}{c}{WD\,J2324}\\
        \hline
         O & \textit{20.2} & \textit{30.3} & \textit{32.6} & \textit{14.5}\\
         Na & 0.06 & 1.04 & \textemdash & \textemdash\\
         Mg & 9.07 & 15.3 & 26.1 & 5.78\\
         Al & 0.41 & 1.70 & 4.03 & 0.83\\
         Si & 13.0 & 21.3 & 15.0 & \textit{6.50}\\
         Ca & 0.60 & 1.61 & 12.2 & 2.24\\
         Fe & 52.6 & 25.9 & 9.19 & 67.4\\
         Ni & 3.04 & 1.64 & \textemdash & 2.75\\
         Other & 1.02 & 1.21 & 0.88 & \textemdash\\
        \hline
    \end{tabular}
\end{table}

WD\,J0358 has a larger fraction of core-like material than bulk Earth, with over 55~per cent of mass as Fe and Ni, and a large abundance of the siderophile Cr. The core mass fraction is similar to other white dwarfs which likely had a small portion of their crust and outer mantle stripped, such as GALEX\,1931+0117 \citep{melis2011accretion} and Ton\,345 \citep{wilson2015composition}. It is also depleted in elements which make up the mantle and crust, with about three times less Ca and almost four times less Al. We also estimate a lower mass fraction of O compared to bulk Earth. Finally, WD\,J0358 has $\mathrm{Ni/Fe}\approx0.05$, similar to the fraction of Fe-Ni alloy in the Earth's core \citep{mcdonough2000} and other white dwarfs accreting core-rich material \citep{williams2025measurements}. This implies that the material accreted onto WD\,J0358 is a core-rich fragment.

WD\,J0426 has mass fractions remarkably similar to bulk Earth, but with slightly more Si and  less Fe. It shows no evidence for the accreted body being a fragment of a differentiated body. However, it is enhanced in the volatile Na which is depleted in the Earth due to volatile loss during heating in the planet formation process and suggests this body formed further out from the host star \citep{wang2019volatility}. This implies a similarity to chondritic material \citep{lodders2025solar}, which further supports the lack of differentiation. Indeed we can see the mass fraction of Na is most similar to that of CI chondrites.

WD\,J0132 is greatly depleted in core-like material, with an Fe fraction closer to that of silicate Earth, but has a much larger amount of Ca even when compared to the continental crust, with a mass fraction of over 12~per cent compared to 4.2~per cent, 2.5~per cent and 1.7~per cent in crust, silicate and bulk Earth, respectively. Pure crust material also has a much smaller fraction of Mg and greater Al and Si. The mass fractions of all refractory elements (Al, Ca, and Ti) are enhanced with respect to both bulk and silicate Earth measurements; it has been shown via simulations that refractory-rich bodies will have formed at high temperatures close in to the host star, within $\sim0.5\,\mathrm{au}$ \citep{bond2010compositional}, although this material would have been engulfed during post-main sequence evolution. However, the composition is not as refractory-dominated as Ca- and Al-rich inclusions (CAIs) which have greater mass fractions of these elements \citep{grossman2008primordial}. The mass fractions may be better represented by a mixture of crust and mantle material.

WD\,J2324 has the largest Fe mass fraction of all four stars and appears to be core-rich rocky material like WD\,J0358, however we cannot make definitive statements based on these mass fractions due to the assumptions used to determine both Si and O abundances. WD\,J0358 has a slighly higher $\log (\mathrm{Mg/Si})=-0.09\pm0.05$ compared to that of the bulk Earth of $\log (\mathrm{Mg/Si})=-0.04$ \citep{mcdonough2000}. Therefore, if we used a similar Mg/Si ratio to WD\,J0358, this would not significantly increase the mass fraction of Si so WD\,J2324 would still be Fe-rich. We conclude that even when considering the systematic uncertainties in the mass fractions of Fe and Ni, WD\,J2324 remains accreting one of the most core-rich fragments, similar to PG\,0843+516 \citep{gansicke2012chemical} and WD\,0059+257 \citep{williams2025measurements}.

Using the mass fractions, we can also quantitatively assess the crust-, mantle-, and core-like nature of the material following the same assumptions as \cite{hollands2018cool}: (1) rocky planetesimals can be described as a linear combination of crust, mantle, and core material, and (2) the abundances of the Earth's crust, mantle, and core are typical for a differentiated planetesimal. As stated in \cite{hollands2018cool}, the first assumption will not hold true for primitive (chondritic) material but would represent the amount of crust, mantle, and core that would have formed if they had undergone differentiation.

We use the crust mass fractions from \cite{lide2004crc} and mantle/core mass fractions from \cite{wang2018elemental}, and combine them to form every permutation of crust/mantle/core material. We then compare these combinations to the mass fractions measured for the four white dwarfs using the $\chi^2$ statistic to find the best match. Our best-fit mixtures are given in \autoref{tab:cmc_frac}.

\begin{table}
    \centering
    \caption{Crust, mantle, and core mass fractions for the four DAZs.}
    \label{tab:cmc_frac}
    \begin{tabular}{lccc}
	\hline
        \multicolumn{1}{l}{Star} & \multicolumn{3}{c}{Mass Fraction ($\pm5$~per cent)}\\
        & \multicolumn{1}{c}{Crust} & \multicolumn{1}{c}{Mantle} & \multicolumn{1}{c}{Core}\\
        \hline
        WD\,J0358 & 0 & 40 & 60\\
        WD\,J0426 & 15 & 60 & 25\\
        WD\,J0132 & 25 & 75 & 0\\
        WD\,J2324 & 5 & 25 & 70\\
        \hline
    \end{tabular}
\end{table}

\subsection{Comparison with Solar System Material}
In order to perform a quantitative comparison of all accreted material compositions to Solar System objects, we perform a goodness-of-fit test (as in \citealt{xu2013two, swan2019interpretation}) by calculating the $\chi^2$ between our measured abundances and those in the Solar System (as previously referenced in \hyperref[sec:abundance_ratios]{Section\,\ref*{sec:abundance_ratios}}). We must convert the absolute abundances into relative abundances to calculate a $\chi^2$ value between the photospheric material and Solar System abundances. This can be written as, e.g.
\begin{equation}
    \mathbf{x}=\begin{pmatrix}
        \log(\mathrm{Mg/H})\\
        \log(\mathrm{Si/H})\\
        \log(\mathrm{Ca/H})\\
        \vdots\\
        \log(\mathrm{Z/H})\\
    \end{pmatrix}.
\end{equation}
We assume that the measurements on each abundance are independent (i.e. the covariance between elements is zero), and so the covariance matrix, $\mathbf{\Sigma}$, is a diagonal matrix, with the diagonal elements equal to the variance, $\sigma^2$ for each element.

Converting the abundances relative to e.g. Ca, we would get
\begin{equation}
    \mathbf{x'}=\begin{pmatrix}
        \log(\mathrm{Mg/Ca})\\
        \log(\mathrm{Si/Ca})\\
        \vdots\\
        \log(\mathrm{Z/Ca})\\
    \end{pmatrix}=\begin{pmatrix}
        1&0&-1&0&\ldots&0\\
        0&1&-1&0&\ldots&0\\
        \vdots&\vdots&\vdots&\vdots&\ddots&\vdots\\
        0&0&-1&0&\ldots&1\\
    \end{pmatrix}\cdot\begin{pmatrix}
        \log(\mathrm{Mg/H})\\
        \log(\mathrm{Si/H})\\
        \log(\mathrm{Ca/H})\\
        \vdots\\
        \log(\mathrm{Z/H})\\
    \end{pmatrix}=\mathbf{Mx},
\end{equation}
where $\mathbf{M}$ is the transformation matrix: $\mathbf{M}$ is essentially an $(N-1)\times(N-1)$ identity matrix with a column of -1 inserted at the position for Ca. The resulting matrix, $\mathbf{x'}$, is an $(N-1)\times N$ matrix that reduces the degrees of freedom by one, which removes the mass of the planetesimal as a free parameter that would otherwise have to be considered.

To ensure $\chi^2$ is independent of the choice of reference element, the covariance matrix must also be transformed according to
\begin{equation}
    \mathbf{\Sigma'}=\mathbf{M\Sigma M}^T.
\end{equation}
We can then calculate the residuals between the abundances and Solar System compositions using
\begin{equation}
    \mathbf{y}=\mathbf{x'}-\mathbf{p},
\end{equation}
with \textbf{p} being the Solar System object in question. The $\chi^2$ statistic is given by
\begin{equation}
    \chi^2=\mathbf{y}^T\mathbf{\Sigma'}^{-1}\mathbf{y}.
\end{equation}

The meteorites are divided into distinct classes and the compositions of these averaged for this analysis, as the properties of material can vary greatly between these groups. For meteorite compositions that were not whole-rock samples (bulk material), these were weighted less in the calculations.

\begin{figure*}
    \includegraphics[width=\textwidth]{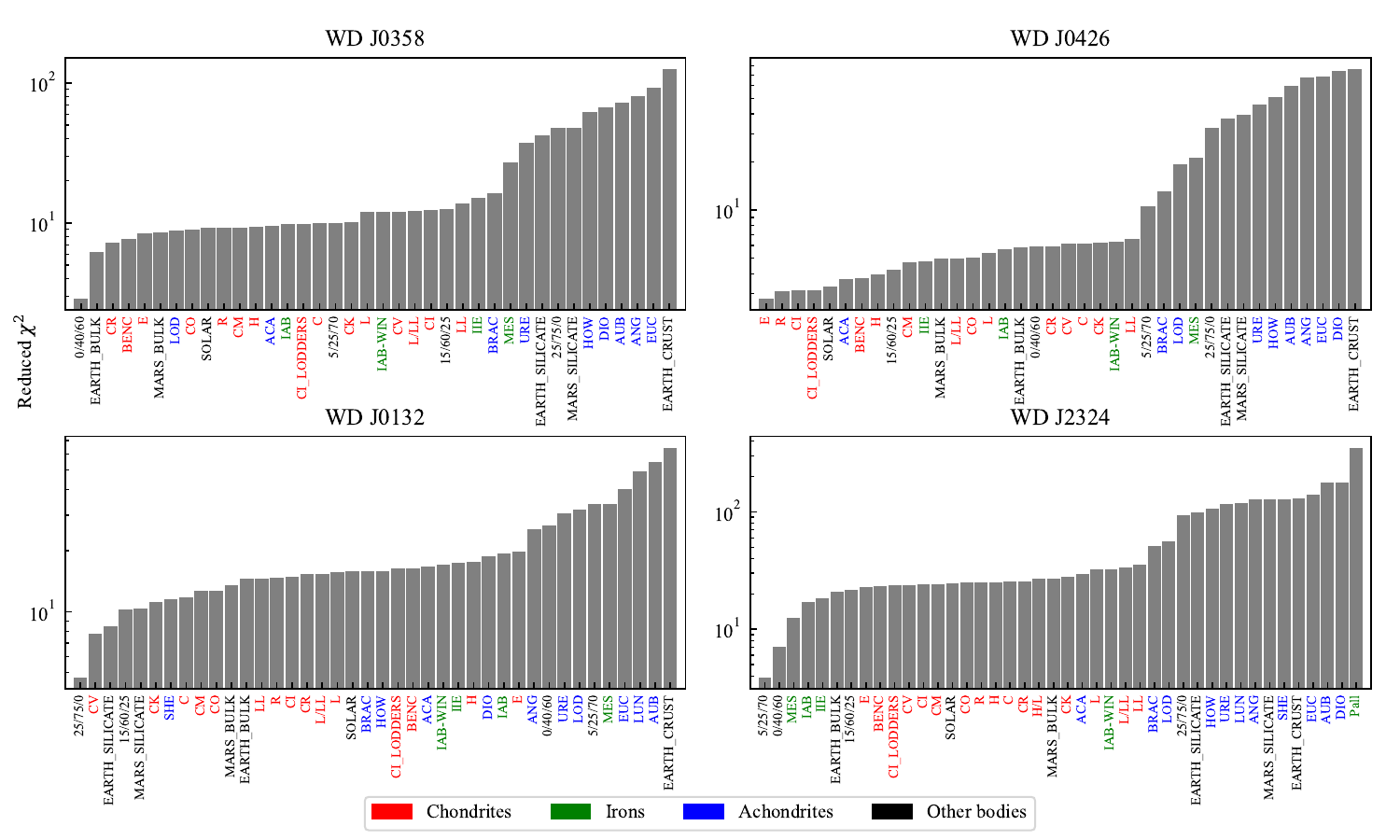}
    \caption{Comparison of Solar System body abundances (Earth, Mars, Solar, Meteorites) to the composition of planetary debris found in the atmosphere of each DAZ in terms of reduced $\chi^2$. We also include the best-fit mixtures of crust/mantle/core calculated in \hyperref[sec:mass_fractions]{Section\,\ref*{sec:mass_fractions}}.}
    \label{fig:chi2}
\end{figure*}

\hyperref[fig:chi2]{Figure\,\ref*{fig:chi2}} displays the results of the comparisons for each DAZ. For three of the DAZs~--~WD\,J0358, WD\,J0132, WD\,J2324~--~we can see there does not seem to be a good match to any meteorite class or other Solar System body. The best matches in each of these cases are in fact the best-fit crust/mantle/core fractions we calculated for each of them in \hyperref[sec:mass_fractions]{Section\,\ref*{sec:mass_fractions}}. This further supports our hypothesis that the planetary material in their photospheres is derived from differentiated bodies. For WD\,J0426, its best-fit crust/mantle/core mixture is not the best match; chondritic meteorites are most similar, which supports our previous statements.

\section{Discussion}
\subsection{Comparison with Cool Polluted White Dwarfs}
\begin{figure}
    \includegraphics[width=\columnwidth]{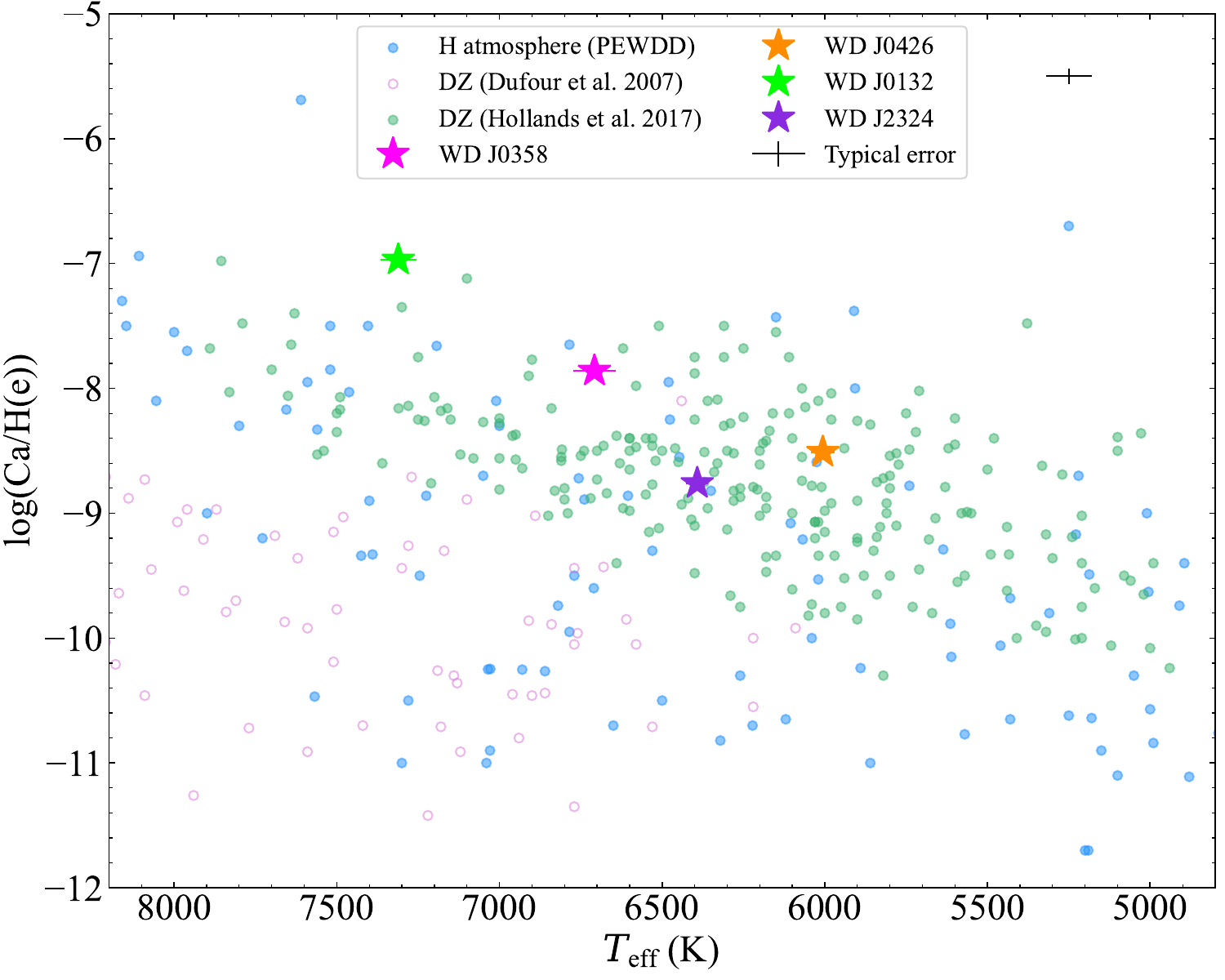}
    \caption{The Ca abundance against effective temperature of the four DAZ white dwarfs from this study compared to all DAZ white dwarfs from PEWDD \citep{williams2024pewdd} and two cool DZ studies \citep{hollands2017cool, dufour2007spectral}.}
    \label{fig:Teff_Ca}
\end{figure}

We now place the four DAZs from this study in the context of other cool polluted white dwarfs using the Planetary Enriched White Dwarf Database (PEWDD, \citealt{williams2024pewdd}). \hyperref[fig:Teff_Ca]{Figure\,\ref*{fig:Teff_Ca}} shows the most common element detected in debris-accreting white dwarfs, Ca, against the effective temperature for all published cool DAZs. This demonstrates that these four are some of the most metal-polluted DAZs for their temperature range, in particular WD\,J0358 and WD\,J0132. Extending this comparison to include He-atmosphere white dwarfs from two large DZ sample studies \citep{dufour2007spectral, hollands2017cool}, we find the Ca abundances in our DAZs are comparable to the most polluted cool DZs, where metals have much longer sinking timescales ($\simeq{10^6}$\,yr) compared to H-atmosphere white dwarfs ($\simeq\mathrm{days}$ to a few 1000\,yr, \citealt{koester2009accretion}). Consequently, DZ white dwarfs are more likely in the decreasing phase, which would cause the measured abundances to differ from the actual composition of the accreted material.

\begin{figure}
    \includegraphics[width=\columnwidth]{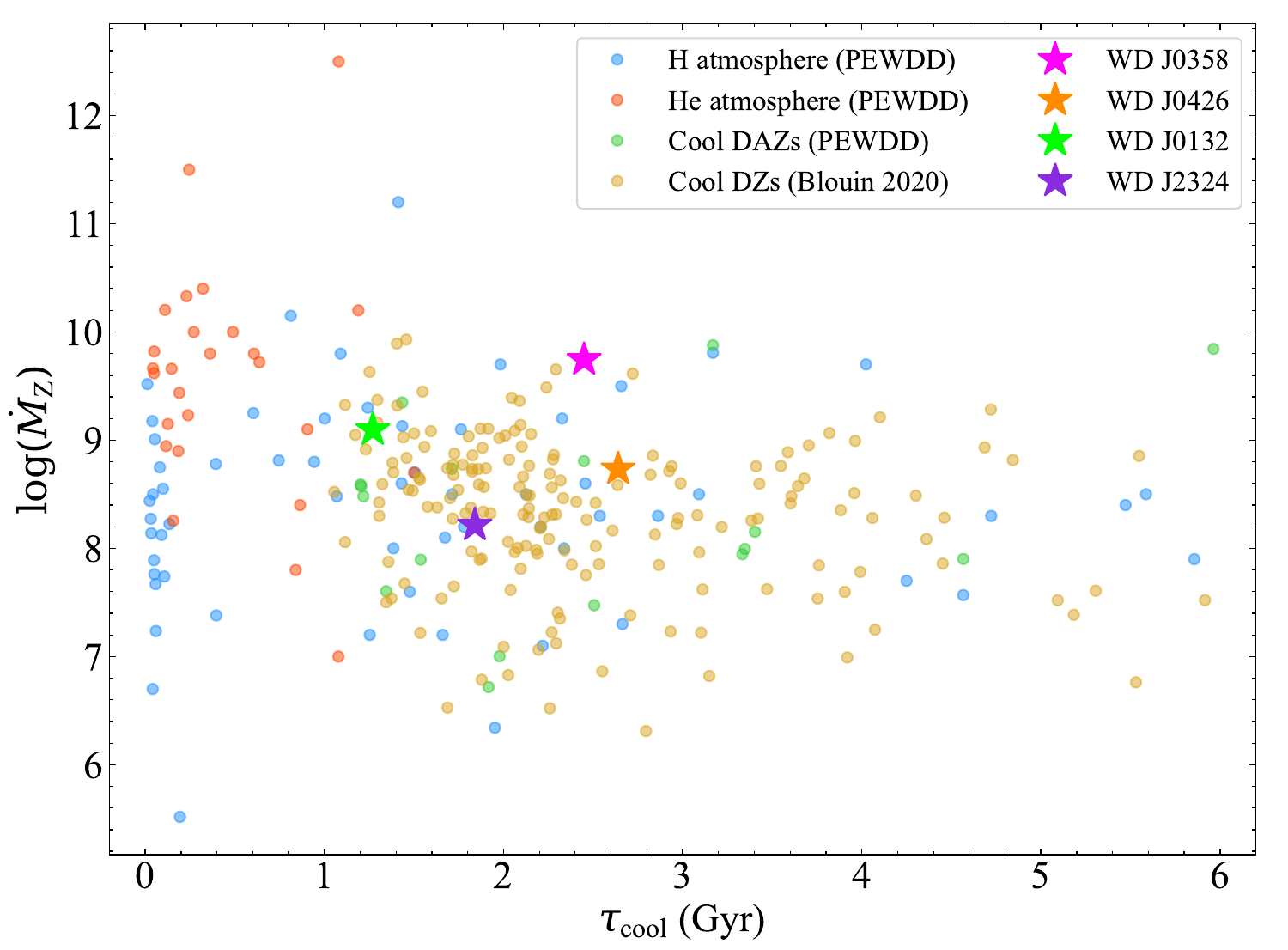}
    \caption{Total accretion rate against cooling age for the four DAZ white dwarfs from this study compared to all stars in PEWDD with published accretion values, labelled as H and He atmosphere, and the cool DAZs in PEWDD and cool DZs in \protect\cite{blouin2020magnesium} with accretion rates inferred from the published metal abundances for each. The cooling age is calculated using the evolutionary tracks of \protect\cite{bedard2020spectral}.}
    \label{fig:Mz_tcool}
\end{figure}

\hyperref[fig:Mz_tcool]{Figure\,\ref*{fig:Mz_tcool}} compares the total accretion rate to the cooling age. All H- and He-atmosphere white dwarfs in PEWDD with published accretion rates are included. \cite{williams2024pewdd} states that there is a dearth of cool He-atmosphere white dwarfs with published accretion rates, so we cannot make any direct comparisons here, but we can see that WD\,J0358 is amongst the highest accretion rates of cool DAZs. We must exercise some caution when interpreting such data, as some studies calculate the total accretion rate by simply scaling up the Ca accretion rate assuming a bulk Earth Ca mass fraction. However, for crust-rich material as in WD\,J0132 this would be an inaccurate assumption; \cite{blouin2022no} use this method for their polluted white dwarf samples and publish a total accretion rate for WD\,J0132 which is about an order of magnitude greater than what we calculate in this work. This may also underpredict accretion rate in core-rich material such as WD\,J0358 which they find to have an accretion rate $\sim0.25\,\mathrm{dex}$ less than in this work.

We can make some assumptions in order to calculate total accretion rates for cool DZs to compare with our sample. We choose the \cite{blouin2020magnesium} sample which is a modified version of the \cite{hollands2017cool} sample with more accurate Mg abundances. We estimate O and Si abundances as $\log(\mathrm{Mg/He})+0.5$ and $\log(\mathrm{Mg/He})+0.04$, respectively, in line with bulk Earth abundances. We follow this approach to obtain values for all of the largest constituents of rocky material, which constitute the dominant contribution to the total accretion rate. The DZ sample has measurements for Ca, Fe and Mg, and so we interpolate the diffusion tables of \cite{koester2020new} to get the sinking times for each of these five metals. We then calculate the accretion rates using \hyperref[eq:7]{Eq.\,\ref*{eq:7}} and sum to get the total accretion rate for each star. We also apply the O and Si estimates to our four stars where needed. The results are also displayed in \hyperref[fig:Mz_tcool]{Fig.\,\ref*{fig:Mz_tcool}}, where we can see our DAZs are comparable to the systems with the highest accretion rates in the cool DZ sample. We also include cool DAZs from PEWDD and apply the same assumptions; however, there are far fewer of these as we apply the condition that they must have at least three detected metals to be included. We must note that this treatment of DZ abundances assumes them to be in steady state, but it is much more likely for them to be in the decreasing phase due to their very long sinking times and cooling ages.

\subsection{Bayesian Analysis}
\begin{figure*}
    \includegraphics[width=\textwidth]{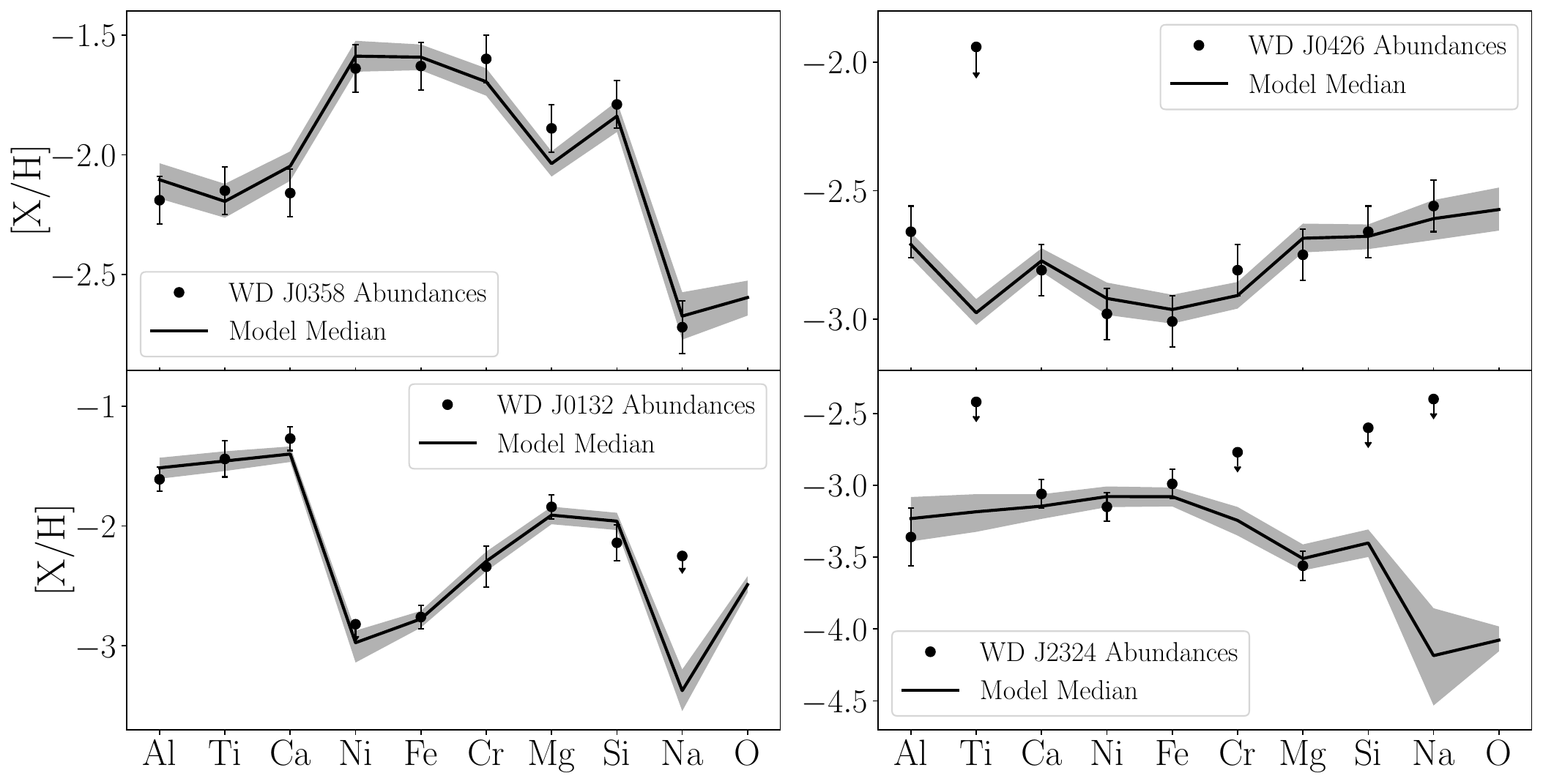}
    \caption{Photospheric metal abundances, and the median fit from the best \textsc{PyllutedWD} model, for each white dwarf. For a given metal X, shown on the horizontal axis, the vertical axis indicates its molar abundance relative to H, normalised to Solar. The grey shaded region indicates the $1\sigma$ confidence interval on the model median. Upper limits are shown with downward pointing arrows. In each case, a (non-constraining) upper limit on O is omitted in order to aid readability. Mn and Co are excluded as these metals are not modelled by \textsc{PyllutedWD}. Top left: For WD\,J0358, the best model invokes preferential accretion of core-like material (with a core mass fraction of $70\pm5$~per cent), due to the high abundances of the siderophile metals Ni, Fe, and Cr. Top right: For WD\,J0426, the accreted material shows no evidence of geological processing. Bottom left: For WD\,J0132, the best model invokes accretion of mantle-like material (with a core mass fraction $3\pm2$~per cent) in order to explain the low Ni, Fe, and Cr abundances/upper bounds. Bottom right: For WD\,J2324, the best model invokes core-like material (with a core mass fraction of $56\pm8$~per cent) due to the high Fe and Ni abundances (relative to Mg). The best model also favours accretion in the decreasing phase in all cases, and invokes incomplete condensation for all cases except WD\,J0426.}
    \label{fig:pylluted_fits}
\end{figure*}

We make use of the Bayesian framework of \cite{buchan2022planets}, \textsc{PyllutedWD}, in order to further analyse our systems. This framework uses the measured metal abundances and white dwarf parameters to find the model that best describes the formation history of the accreted body. It is assumed that the accreted material is, in general, a fragment of a planetary body which formed from the same stellar nebula as the white dwarf progenitor. Models are constructed by invoking, or omitting, key geological processes. The two most important processes are incomplete condensation (depletion of volatile metals due to formation close to the host star), and core-mantle differentiation (formation of an Fe-rich core and an Fe-poor mantle). The framework favours simpler models with fewer parameters (i.e., those which invoke the fewest geological processes). All models take into account the differential sinking of elements in the white dwarf's atmosphere to trace the measured abundances back to the composition of the accreted material. They also predict the phase of accretion we observe and the duration of the accretion event. The resulting model fits are shown in \hyperref[fig:pylluted_fits]{Fig.\,\ref*{fig:pylluted_fits}}.

Due to the small statistical errors on the calculated abundances, we adopt more conservative errors for this analysis in line with the literature by setting the minimum abundance error to 0.1\,dex.

For WD\,J0426, the best model describes the pollutant as primitive material with no geological processes modifying the composition from that of the initial stellar nebula. The total accreted mass is predicted to be $10^{21.5\pm 0.2}\,\mathrm{g}$. The best model favours high metallicity in the stellar nebula.

For the other three of our DAZs (WD\,J0358, WD\,J0132, WD\,J2324), the model with the highest Bayesian evidence invokes core-mantle differentiation to $5.7\sigma$, $4.1\sigma$, and $2.3\sigma$ confidence respectively, and incomplete condensation in the parent body to $7.4\sigma$, $14.0\sigma$, and $2.3\sigma$ confidence respectively. 

The material accreted onto two of these~--~WD\,J0358 and WD\,J2324~--~are from core-rich fragments with masses of order $10^{21}\,\mathrm{g}$ and $10^{19}\,\mathrm{g}$, and core mass fractions of $70\pm5$~per cent and $56\pm8$~per cent, respectively. The third~--~WD\,J0132~--~has accreted a fragment of nearly pure mantle material with a mass of order $10^{20}\,\mathrm{g}$ and core mass fraction of just $3\pm2$~per cent.

These results are broadly consistent with those found in \hyperref[sec:mass_fractions]{Section\,\ref*{sec:mass_fractions}}. However, while those results implied a higher core mass fraction for WD\,J2324 than for WD\,J0358, here we find that the opposite is true. The primary factor is that, for WD\,J0358, Al, Ti and Ca are depleted relative to chondritic material (as shown in \hyperref[fig:pylluted_fits]{Fig.\,\ref*{fig:pylluted_fits}}). This pushes \textsc{PyllutedWD} more strongly towards core-rich solutions in this case. In \hyperref[sec:mass_fractions]{Section\,\ref*{sec:mass_fractions}}, these metals have only a small effect on the mass fractions of major elements because they are naturally less abundant in chondritic material.

\subsubsection{Accretion Phase}

In all four systems, the best model infers that the accretion event has already ended, and the system has reached the decreasing phase of accretion. The probability of being in the decreasing phase is at least 88~per cent in all cases. This result may be unexpected, as H-atmosphere white dwarfs are often assumed to be in the steady state of ongoing accretion on the basis that their metal sinking timescales are short relative to the typical duration of accretion events. However, since these DAZs are cool, their sinking timescales are significantly increased such that this condition might no longer hold, depending on the assumptions made about accretion event duration. The sinking timescales for these DAZs are of order $10^3-10^4$\,yr which is shorter than disc lifetimes as predicted from dynamics ($\sim10^5$\,yr, \citealt{rafikov2011runaway}) or as estimated from observational constraints ($10^{6.1\pm1.4}$\,yr,  \citealt{cunningham2021horizontal}). It is, however, significantly longer than accretion lifetime estimates based on the statistics of metal-polluted white dwarfs ($\simeq20$\,yr, \citealt{wyatt2014stochastic}).

The prior distribution on the accretion event lifetime used by \textsc{PyllutedWD} is broad, ranging from $10^0$ to $10^8$\,yr. Given that the lower limit is significantly shorter than the sinking timescales for these white dwarfs, decreasing phase solutions are admissible. By modifying this prior distribution, we verified that steady state solutions can be found for all systems (and that therefore the preference for decreasing phase solutions is driven by our prior assumptions). Forcing the solution into the steady state phase has no qualitative effect on other results, except that for WD\,J0426 the best model now invokes weak evidence of incomplete condensation (to $1.1\sigma$ significance). The prior distribution used here contrasts with the adoption of steady state accretion used elsewhere in this paper, which implicitly assumes accretion event lifetimes $\gg10^4$\,yr.

Even though we can attribute the decreasing-phase results to our prior assumptions, it is instructive to consider what the physical interpretation would be. It would be surprising to have caught all four stars at a special time in their history, just after the end of a large accretion event. A simpler interpretation would be that our model of a constant-rate accretion episode is wrong, and that the stars show evidence for variability in their accretion rates on timescales of $\lesssim10^4$\,yr. Time-averaged accretion rates appear to differ between H- and He-atmosphere white dwarfs \citep{farihi2012scars}, potentially due to temporarily enhanced rates of accretion linked to volatiles \citep{okuya2023modelling}, and accretion rate changes have been observed \citep{farihi2026accretion}, so our results would be consistent with that picture. These considerations motivate further analysis of metal-polluted white dwarfs whose sinking timescales lie in the intermediate range of accretion event duration estimates.

\subsubsection{Abundance Errors}

We assume a (minimum) error of 0.1\,dex on metal abundances, which is lower than typical literature values \citep{williams2024pewdd}. We re-ran the Bayesian framework using a minimum error of 0.2\,dex, finding that in most cases there is no qualitative change to our results (although, as expected, their significance decreases). However, for WD\,J2324 the best model no longer invokes differentiation or incomplete condensation, favouring accretion of primitive material instead. Of the three systems which showed evidence for geological processing, the evidence was weakest in this case ($2.3\sigma$).

\subsection{Infrared Excess}
The SEDs in \hyperref[fig:sed]{Fig.\,\ref*{fig:sed}} show a slight infrared (IR) excess in the \textit{WISE} W2 band for WD\,J0358 and WD\,J2324 which could potentially signify the presence of a debris disc. Including a conservative five~per cent systematic uncertainty alongside the quoted W2 error, we find the excess to be $5.0\sigma$ for WD\,J0358 and $2.2\sigma$ for WD\,J2324.  We fit a white dwarf + disc model to the near-IR data to determine the inner disc temperature and radius of these hypothetical discs. For the white dwarf model, we simply use a Planck function, given by
\begin{equation}
    B_{\nu} = \frac{2h\nu^3}{c^2}\frac{1}{e^\frac{h\nu}{k_\mathrm{B}T}-1}.
\end{equation}
We define the disc using the equations from \cite{jura2003tidally}. This model has been shown to have significant limitations in accounting for the variety of dusty white dwarf observations \citep{farihi2018dust, gentile2019gaia, swan2020dust, noor2025activity}. However, we use it to facilitate comparison against other discs fitted with the same model. 

The temperature, $T_\mathrm{ring}$ as a function of distance from the star, $R$, is given by
\begin{equation}
    T_\mathrm{ring}\approx\left(\frac{2}{3\pi}\right)^\frac{1}{4}\left(\frac{R_*}{R}\right)^\frac{3}{4}T_*\label{eq:19},
\end{equation}
and the flux from the ring, $F_\mathrm{ring}$, is given by
\begin{equation}
    F_\mathrm{ring}=\frac{2\pi\cos{i}}{D_*^2}\int_{R_\mathrm{in}}^{R_\mathrm{ou}}B_\nu(T_\mathrm{ring})R\,dR.
\end{equation}

\begin{figure*}
\centering
\begin{subfigure}{0.49\textwidth}
    \includegraphics[width=\columnwidth]{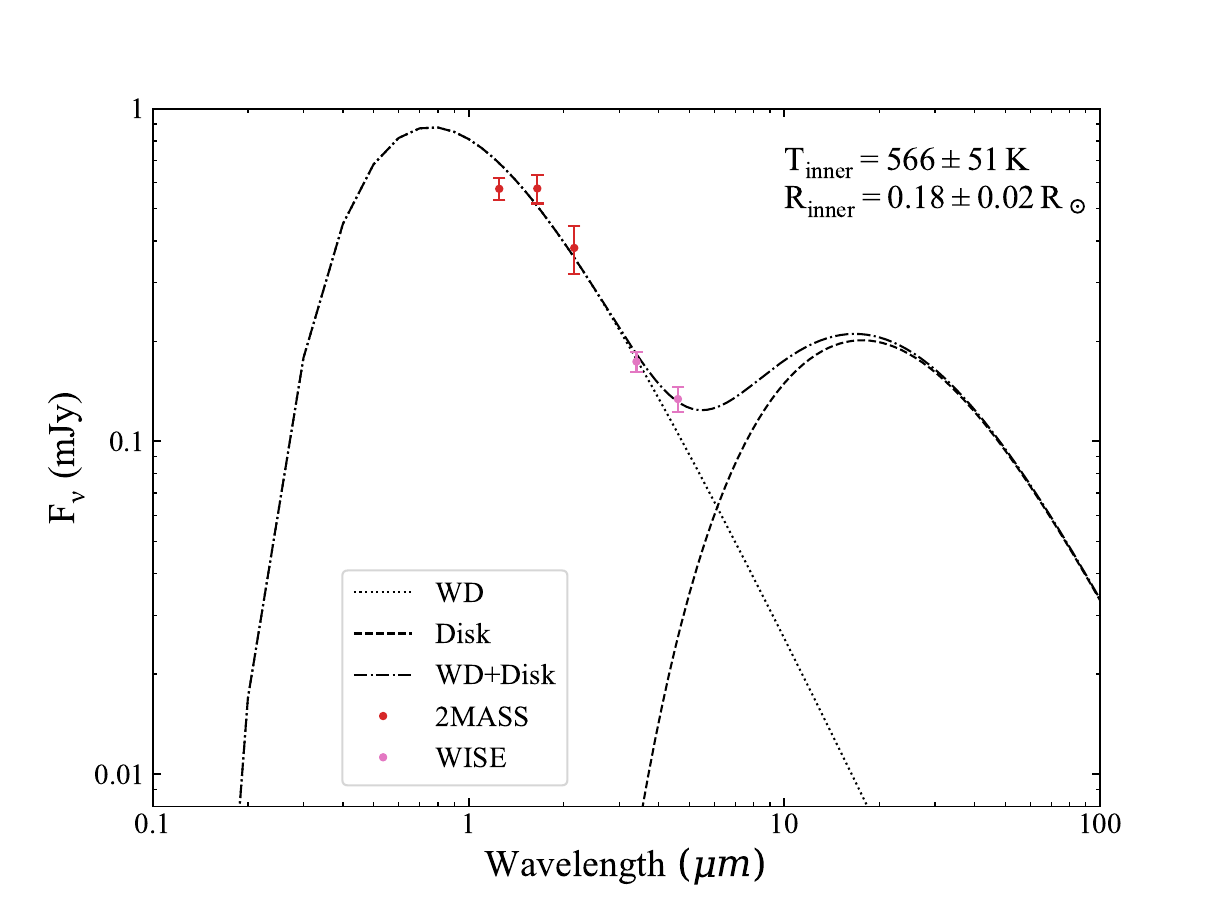}
    \caption{WD\,J0358}
    \label{fig:disk_wdj0358}
\end{subfigure}
\hfill
\begin{subfigure}{0.49\textwidth}
    \includegraphics[width=\columnwidth]{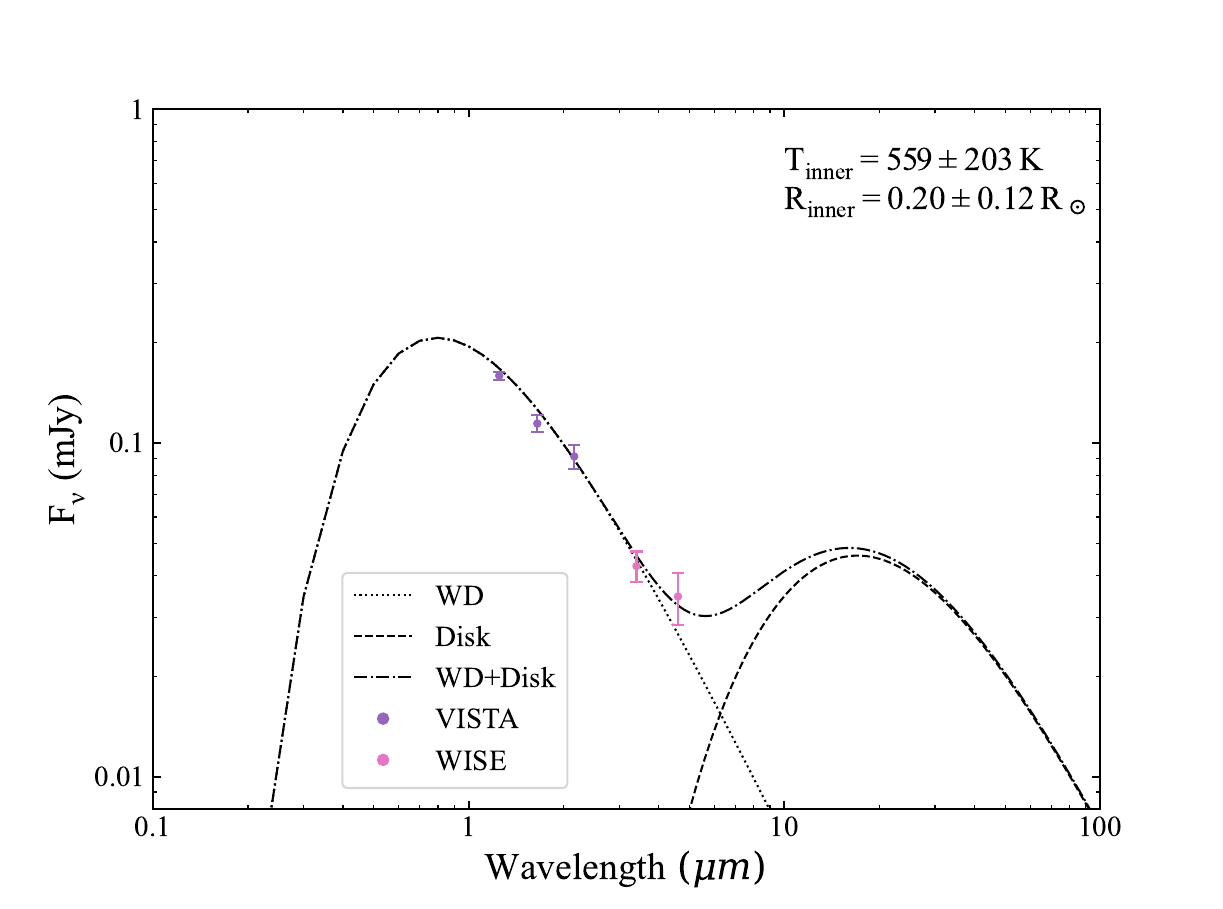}
    \caption{WD\,J2324}
    \label{fig:disk_wdj2324}
\end{subfigure}
\caption{Best-fit disc models for near-IR photometry.}
\end{figure*}

We fix the inclination to 52 degrees as this has been found to be the average inclination of observed discs\footnote{https://keatonb.github.io/archivers/uniforminclination}, and set $R_\mathrm{out}$ to 1$R_\odot$ (which in turn provides $T_\mathrm{out}$ using \hyperref[eq:19]{Eq.\,\ref*{eq:19}}). The tidal disruption radius provides the limit on $R_\mathrm{out}$, as we lack data in the further-IR to constrain it better. We then use $\chi^2$ minimisation to find the best-fitting disc parameters, with $T_\mathrm{in}$ as the free parameter~--~limited to a maximum of 1200\,K where silicate grains sublimate~--~from which we can subsequently calculate $R_\mathrm{in}$. \hyperref[fig:disk_wdj0358]{Figure\,\ref*{fig:disk_wdj0358}} shows the best model for WD\,J0358 with $T_\mathrm{in}=566$\,K and $R_\mathrm{in}=0.18$\,$R_\odot$, and \hyperref[fig:disk_wdj2324]{Fig.\,\ref*{fig:disk_wdj2324}} shows the best model for WD\,J2324 with $T_\mathrm{in}=559$\,K and $R_\mathrm{in}=0.20\,R_\odot$. These are cool compared to disc observations \citep{rocchetto2015frequency}.

\textit{WISE} has been shown to have insufficient spatial resolution \citep{dennihy2020word}. In contrast, newer IR missions such as \textit{JWST} have the capabilities of confirming and characterising these tentative IR excess detections, being superior to \textit{WISE} in sensitivity, resolution, and wavelength coverage.

\section{Conclusions}
This paper presents the analysis of four cool and highly metal-polluted DAZ white dwarfs: WD\,J0358, WD\,J0426, WD\,J0132, and WD\,J2324. We fit model grids to the photometric data of each star, and obtained the photospheric parameters of $T_\mathrm{eff}$ and $\log g$. We also fit these grids to the X-shooter spectra to measure the metal abundances of the accreted material and infer the parent body compositions. 

We identified eleven metals in WD\,J0358, ten metals in WD\,J0426, seven metals in WD\,J0132, and five metals in WD\,J2324. There are currently only eight cool H-dominated white dwarfs with five or more metals according to PEWDD, and so this study increases the sample by 50~per cent. We used the measured abundances and accretion rates to calculate mass fractions and compared the accreted material compositions to meteorites and other Solar System bodies. 

We find that the planetesimal accreted by WD\,J0426 is likely composed of primitive chondritic material with no geological modification, but the debris accreted by WD\,J0358, WD\,J0132, and WD\,J2324 all show evidence of being from differentiated bodies. WD\,J0358 and WD\,J2324 are both core-rich with Fe mass fractions of 53~per cent and 67~per cent respectively, whilst WD\,J0132 is mantle-rich with a Ca mass fraction of 12~per cent.

We conclude that these are some of the most heavily polluted cool DAZs studied to date with high accretion rates suggesting ongoing accretion, although uncertainties in accretion lifetime could mean they may have very recently stopped accreting. As such, this is the first equivalent to the large sample of strongly polluted cool He-atmosphere white dwarfs (e.g. \citealt{hollands2017cool}), which have much longer diffusion timescales compared to H atmospheres and so have most likely finished the accretion episode. Future multi-object spectroscopic surveys such as DESI, WEAVE, and 4MOST will discover more cool DAZs amenable to chemical characterisation of the debris.

\section*{Acknowledgements}
We thank the anonymous referee for their thorough and insightful report. AMC acknowledges support from grant project reference 2590460 from the Science and Technology Facilities Council (STFC). This project has received funding from the European Research Council (ERC) under the European Union’s Horizon 2020 research and innovation programme (Grant agreement No. 101020057 (BTG, AS, PI, JTW) and 101002408 (AMB)). This paper is based on observations collected at the European Organisation for Astronomical Research in the Southern Hemisphere under ESO programmes 1103.D-0763(C), 093.C-0275(A), and 0101.C-0646(A). This work has made use of data from the ESA mission Gaia (\url{https://www.cosmos.esa.int/gaia}), processed by the Gaia Data Processing and Analysis Consortium (DPAC, \url{https://www.cosmos.esa.int/web/gaia/dpac/consortium}). Funding for the DPAC has been provided by national institutions, in particular the institutions participating in the Gaia Multilateral Agreement. This work made use of \textsc{Astropy}\footnote{http://www.astropy.org}: a community-developed core Python package and an ecosystem of tools and resources for astronomy \citep{price2022astropy}, and \textsc{scipy} \citep{virtanen2020scipy}. This research has made use of the VizieR catalogue access tool, CDS, Strasbourg Astronomical Observatory, France \citep{ochsenbein2000vizier}.
\section*{Data Availability}
The X-shooter data used in this paper is available in the ESO archives under programme IDs 1103.D-0763(C), 093.C-0275(A), and 0101.C-0646(A).



\bibliographystyle{mnras}
\bibliography{references}



\appendix
\section{Photometric data}
\begin{table*}
	\centering
	\caption{Photometric data for the four white dwarfs. \textit{GALEX} and \textit{WISE} points are not used in the fitting for reasons described in \hyperref[sec:phot_fits]{Section\,\ref*{sec:phot_fits}}.}
	\label{tab:phot_table}
	\begin{tabular}{@{}llllll}
		\hline
		\multicolumn{1}{l}{Survey} & \multicolumn{1}{l}{Filter} & \multicolumn{4}{c}{Magnitude} \\
		& & \multicolumn{4}{c}{(mag)} \\
            & & \multicolumn{1}{c}{\textbf{WD\,J0358}} & \multicolumn{1}{c}{\textbf{WD\,J0426}} & \multicolumn{1}{c}{\textbf{WD\,J0132}} & \multicolumn{1}{c}{\textbf{WD\,J2324}} \\
            \hline
            \multirow{3}{*}{\hspace{-0.1cm}\minitab{\textit{Gaia} DR3 \\ \cite{gaia2023dr3}}} & $G$ & 16.646$\pm$0.003 & 16.793$\pm$0.003 & 18.208$\pm$0.003 & 18.279$\pm$0.003 \\
		  & $G_\mathrm{BP}$ & 16.862$\pm$0.007 & 17.080$\pm$0.009 & 18.407$\pm$0.014 & 18.497$\pm$0.033 \\
		& $G_\mathrm{RP}$ & 16.294$\pm$0.007 & 16.351$\pm$0.006 & 17.913$\pm$0.018 & 17.871$\pm$0.029 \\
		\hline
		\multirow{5}{*}{\hspace{-0.1cm}\minitab{Pan-STARRS \\ \cite{chambers2016pan}}} & \textit{g} & 16.868$\pm$0.019 & \multirow{5}{*}{\textemdash} & 18.363$\pm$0.004 & 18.529$\pm$0.007 \\
		& \textit{r} & 16.675$\pm$0.010 & & 18.251$\pm$0.005 & 18.293$\pm$0.005 \\
		& \textit{i} & 16.621$\pm$0.010 & & 18.235$\pm$0.005 & 18.219$\pm$0.004 \\
		& \textit{z} & 16.653$\pm$0.009 & & 18.295$\pm$0.005 & 18.221$\pm$0.005 \\
		& \textit{y} & 16.684$\pm$0.034 & & 18.360$\pm$0.026 & 18.240$\pm$0.007 \\
		\hline
            \multirow{5}{*}{\hspace{-0.1cm}\minitab{SDSS \\ \cite{ahn2014tenth}}} & \textit{u} & \multirow{5}{*}{\textemdash} & \multirow{5}{*}{\textemdash} & 18.946$\pm$0.023 & 19.190$\pm$0.027 \\
            & \textit{g} & & & 18.385$\pm$0.007 & 18.571$\pm$0.008 \\
            & \textit{r} & & & 18.237$\pm$0.008 & 18.290$\pm$0.008 \\
            & \textit{i} & & & 18.212$\pm$0.010 & 18.214$\pm$0.009 \\
            & \textit{z} & & & 18.230$\pm$0.025 & 18.199$\pm$0.027 \\
            \hline
		\multirow{3}{*}{\hspace{-0.1cm}\minitab{2MASS \\ \cite{cutri20032mass}}} & J & 16.108$\pm$0.081 & 15.856$\pm$0.064 & \multirow{3}{*}{\textemdash} & \multirow{3}{*}{\textemdash} \\
		& H & 15.625$\pm$0.107 & 15.700$\pm$0.116 & & \\
		& Ks & 15.606$\pm$0.179 & 15.617$\pm$0.216 & & \\
		\hline
		\multirow{5}{*}{\hspace{-0.1cm}\minitab{UKIDSS \\ \cite{lawrence2007ukirt}}} & Z & 16.121$\pm$0.006 & \multirow{5}{*}{\textemdash} & & \multirow{5}{*}{\textemdash} \\
		& Y & 16.163$\pm$0.007 & & 17.842$\pm$0.025 & \\
		& J & 15.966$\pm$0.009 & & 17.758$\pm$0.040 & \\
		& H & 15.733$\pm$0.016 & & 17.403$\pm$0.061 & \\
		& K & 15.749$\pm$0.014 & & 17.518$\pm$0.120 & \\
		\hline
            \multirow{6}{*}{\hspace{-0.1cm}\minitab{SkyMapper \\ \cite{wolf2018skymapper}}} & $\textit{u}$ & \multirow{6}{*}{\textemdash} & 18.081$\pm$0.023 & \multirow{6}{*}{\textemdash} & \\
            & \textit{v} & & 17.785$\pm$0.023 & & \\
            & \textit{g} & & 17.060$\pm$0.019 & & 18.450$\pm$0.046 \\
            & \textit{r} & & 16.764$\pm$0.007 & & 18.291$\pm$0.062 \\
            & \textit{i} & & 16.706$\pm$0.043 & & 18.253$\pm$0.031 \\
            & \textit{z} & & 16.747$\pm$0.051 & & 18.422$\pm$0.105 \\
            \hline
            \multirow{4}{*}{\hspace{-0.1cm}\minitab{VISTA \\ \cite{mcmahon2013first}}} & Y & \multirow{4}{*}{\textemdash} & & \multirow{4}{*}{\textemdash} & 17.628$\pm$0.028 \\
            & J & & 15.891$\pm$0.005 & & 17.461$\pm$0.029 \\
            & H & & & & 17.372$\pm$0.064 \\
            & Ks & & 15.599$\pm$0.015 & & 17.149$\pm$0.090 \\
            \hline
		\vspace{0.4cm}\multirow{2}{*}{\hspace{-0.1cm}\minitab{\textit{GALEX} \\ \cite{bianchi2017revised}}} & \multirow{2}{*}{NUV} & \multirow{2}{*}{19.787$\pm$0.136} & \multirow{2}{*}{20.934$\pm$0.236} & \multirow{2}{*}{20.387$\pm$0.125} & \multirow{2}{*}{21.692$\pm$0.250} \\
		\hline
		\multirow{2}{*}{\hspace{-0.1cm}\minitab{\textit{WISE} \\ \cite{marocco2021catwise2020}}} & W1 & 15.627$\pm$0.023 & 15.581$\pm$0.019 & 17.532$\pm$0.087 & 17.150$\pm$0.061 \\
		& W2 & 15.269$\pm$0.037 & 15.751$\pm$0.043 & & 16.738$\pm$0.140 \\
		\hline
	\end{tabular}
\end{table*}

\section{Metal absorption line wavelengths}
\begin{table}
    \centering
    \caption{Vacuum wavelengths of the strong metal lines found in WD\,J0358; Fe lines were too numerous to list individually. A subset of these lines are present in each of the other three DAZs.}
    \label{tab:strong_lines}
    \begin{tabular}{p{0.1\linewidth}p{0.8\linewidth}}
        \hline
        Ion & Vacuum Wavelength (\AA)\\
        \hline
        \ion{Na}{I} & 5891.583, 5897.558\\
        \ion{Mg}{I} & 3830.441, 3833.391, 3839.381, 3879.405, 3896.676, 3987.881, 5168.761, 5174.125, 5185.048\\
        \ion{Al}{I} & 3945.122, 3962.641\\ 
        \ion{Si}{I} & 3906.629\\
        \ion{Ca}{I} & 4227.918\\
        \ion{Ca}{II} & 3934.778, 3969.592, 8500.355, 8544.435, 8664.518\\
        \ion{Ti}{II} & 3235.448$^*$, 3237.506$^*$, 3239.971$^*$, 3242.918, 3342.835, 3350.365, 3362.178, 3373.761, 3384.73, 3388.806\\
        \ion{Cr}{I} & 3369.009, 3409.735, 3422.183, 3423.714, 3579.708, 3594.511, 3606.358, 4255.534, 4276.015, 4290.924, 5205.960, 5207.487, 5209.875\\
        \ion{Mn}{I} & 4031.892, 4034.202, 4035.623\\
        \ion{Fe}{I} & $\sim200$ lines between 3190-5460\\
        \ion{Co}{I} & 3406.091$^*$, 3410.151, 3413.312, 3450.156, 3454.497$^\ddagger$, 3475.035, 3484.402$^\ddagger$, 3503.280, 3507.315, 3510.844$^\ddagger$, 3513.642$^*$, 3530.815, 3846.552, 3895.177\\
        \ion{Ni}{I} & 3225.947$^*$, 3233.863, 3235.577, 3243.988, 3362.517, 3367.129, 3370.531, 3372.951, 3375.184, 3381.537, 3392.012, 3393.957$^*$, 3415.740$^*$, 3424.686$^*$, 3434.538, 3438.260, 3447.242, 3453.875$^\dagger$, 3459.447, 3462.641, 3473.536, 3484.767$^\dagger$, 3493.954, 3501.848, 3511.336$^\dagger$, 3516.054, 3520.765, 3525.542, 3549.193, 3567.385$^*$, 3572.880$^*$, 3598.727, 3611.491$^*$, 3613.763, 3620.418$^*$, 3675.185, 3776.637, 3784.599, 3808.219, 3859.386$^*$\\
    \end{tabular}
    \footnotesize{$^*$Blended with \ion{Fe}{I}, $^\dagger$Blended with \ion{Co}{I}, $^\ddagger$Blended with \ion{Ni}{I}}
\end{table}


\bsp	
\label{lastpage}
\end{document}